# 湯川秀樹先生のはじめての胸像は何故高知に建てられたか
# First bronze statue of Prof. Hideki Yukawa in Kochi

大阪大学核物理研究センター
Research Center for Nuclear Physics, Osaka University, Ibaraki 567-0047, Japan
大久保茂男*i
Shigeo Ohkubo

要約　我が国初のノーベル物理学賞受賞者で中間子論の湯川秀樹博士の胸像（浜口青果作）が 1954 年 3 月高知県夜須町の小学校の校庭に、5 年間の米国滞在から帰国してまもない湯川秀樹博士および澄子夫人が出席して除幕式が行われ建立されていたことがわかった。この我が国最初の湯川胸像は、よく知られている京都大学基礎物理学研究所湯川記念館前に湯川財団により設置された胸像（1986 年山本挌二作）より 32 年以上前に、夜須小学校 PTA の住民運動により設置されていた。何故高知の夜須小学校に湯川胸像が最初に建てられ、なぜ湯川夫妻が出席したのか、その経緯・背景が明らかにされる。

Abstract: The first bronze statue of Prof. Hideki Yukawa, the first Nobel Prize laureate in Japan, is found to have been built in March 1954 in the Yasu elementary school in the countryside of Kochi prefecture in Shikoku Island. It is also found that the bust sculpture was unveiled with the attendance of Prof. Yukawa and his wife, who were invited by the local people, soon after their coming back to Japan from USA where they stayed for five years for research. This bust bronze statue was contributed by the local people of the elementary school more than 32 years earlier than the well-known one of Prof. Yukawa built in 1986 by the Yukawa foundation in front of the Yukawa Hall, Yukawa Institute for Theoretical Physics, Kyoto University in Kyoto city. Why the statue was built in Kochi is unveiled and the historical background is discussed.

## １．はじめに

我が国初のノーベル物理学賞受賞者である湯川秀樹先生(1907-1981 年)の随筆集に『しばしの幸』[1]がある。読売新聞社に依頼され、5 年間の米国滞在から帰国後、1953 年秋から翌年はじめまでに書かれたものである。「はしがき」によると、「去年の夏の暑いさかりに帰ってきてから、引続き身辺多事で、六年ぶりで京の桜の時節を迎えることになったのに・・・」とある。最初の随筆は「京の秋」である。冒頭の書き出しは「国際理論物理学会議が終わつたと思うと台風という歓迎されないお客様が訪れた。それがすんだら急に秋が深くなつてきた。私にとつて日本の秋は六年ぶりである。夏休みを利用してアメリカか

---

*Research fellow of RCNP, Emeritus Prof. of Univ. of Kochi, 高知県立大学名誉教授

ら帰つてきても、九月の声をきくと渡り鳥のように飛び去つてゆかねばならなかつた。今年はじめてゆつくり京都の秋が満喫できるのである。」で始まる味わいのある文である。この随筆には10番目の「北海道の夏」に続いて「四国の秋」[1][2]がある。中間子論[3]を1935年に発表していたが、まだ世に知られてない1938年大阪大学助教授時代の湯川先生が、徳島の中学の校長に初めての一般むけの講演を依頼され、家族総ぞろいで徳島に出かけていった時の様子が書かれている。1938年の秋、徳島中学のクラスの授業で最新の中間子論[3]にもふれた物理学の話をしたものである。いまでいうアウトリーチの先駆けである。北海道とともに「まだ見ぬ四国に何となく魅力を感じていた」[1][2]湯川先生はこの四国訪問でうけた大変強い印象をつづっている。私は数年前、湯川自身の監修をへたと思われるこの時の講演記録が京都大学基礎物理学研究所の湯川記念資料室の湯川の所持品の中にあるのではないかと予感し発見した[4]。湯川が目を通した証拠に、見つかった講演記録には自身でつけた赤鉛筆の丸印が残されていた。70年以上前のわかわかしい物理学者湯川に会ったような気がした。このことを雑誌「素粒子論研究電子版」に「湯川秀樹先生の初めての一般講演」と題して紹介した[4]。

　湯川先生はその後徳島以外の四国、とくに高知を訪れたことがあるだろうか。高知市生まれの私はそのことに関心があったが、湯川秀樹著作集別巻の年譜[5]にも高知訪問は載っていない。上記の「湯川秀樹先生の初めての一般講演」の調査で2013年2月に基礎物理学研究所の湯川記念館史料室の資料を当時の記念館史料室責任者、九後太一基礎物理学研究所長とくまなく調べおりにも、記録はみあたらなかった。

　私の専門である原子核理論の「原子核虹」の研究[6]（筆者は湯川虹†とよぶことを提唱している）、「湯川核力の引力芯説」の提唱[8]、および「場の理論的クラスター模型による原子核構造」の研究[9]などに没頭し、しばらく忘れかけていたが、2018年になって故郷の高知で古い記録を探してみると、湯川先生が高知を1954年春に訪れていることがわかった。「四国の春」である。残念ながら湯川先生の「四国の春」という随筆は残されてない。講演会なら湯川先生は日本各地を訪問されているので、高知市に来ていてもさほど驚くことでもない。驚愕したのは単なる講演のためではなく、日本で初めての「湯川胸像」除幕式に出席するためであることがわかったことである。しかも胸像は高知市ではなく、はるか離れた、田舎の小学校に設立されていた。なぜ、地方の小学校に湯川先生の胸像が設立され、湯川先生夫妻が出席されたのか。私は京都大学で3回生の時に湯川先生の講義をうけ直接ならい、また大学院でも影響をうけたのでいっそう関心が強くなった。

　湯川先生のノーベル賞受賞の中間子論[3]にかんする記念像といえば、中間子論の研究を行ったときの西宮市苦楽園の住居に近い苦楽園小学校の校庭に「中間子論誕生記念碑」が1985年11月2日に除幕式が行われ建設されたことがよく知られている[10]。しかしここには湯川の銅像は建てられてない。湯川先生の胸像は京都大学基礎物理学研究所の湯川記念

---

† 気象虹は「ニュートン虹」とよばれ、湯川核力で起こる原子核虹はユカワ虹とよぶことができる[6][7]。

館前にあり(図1)、一般にもひろく知られている。筆者が京都大学の学生・大学院生であった1960、1970年代には胸像はなかった。胸像裏面の設立趣意書によると、1986年12月湯川財団から京都大学へ寄贈され設立されたとあり、設置主体は京都大学である。『京都大学百年史』[11]は設置のいきさつを次のように記している。「昭和56(1981)年の湯川の死後、湯川と京都一中で同窓生だった彫刻家菊池一雄は、湯川の胸像制作を構想したが、同60年死去した。財団は湯浅(湯川記念財団理事長のこと、筆者)から胸像制作費として指定寄付を受け、湯川の京都一中の後輩である彫刻家山本挌二(京都市立芸術大学名誉教授)に制作を依頼した。胸像は基研(基礎物理学研究所のこと、筆者)湯川記念館前の木立の中に設置され、昭和61(1986)年12月11日に除幕式が行われた。」

　図1の湯川胸像写真は筆者が2018年初夏に京都大学基礎物理学研究所で開かれた長期間の国際研究集会「QCDの新しい展開　2018」(5月28日－6月29)[12]に参加したおりに撮影したものである。胸像は以前は『京都大学百年史』に記載されているように木立の中にあって、近くまで行けなかったが、図1(a)のように、植え込みに参道のように切り込道が設けられ、まぢかで見られるようになっている。修学旅行などで訪れる生徒をときどき見かけることがあり、人気の場所のようだ。かつて賽銭がおかれているのをみかけたが、受験生や生徒にはそういう心情が働くのだろうか。今回は賽銭を見かけなかったが、かつて湯川記念館ができたとき、湯川先生も加わった草創期1957年10月の座談会‡で湯川記念館は＜メッカか神社か＞[13]という議論があったのを思い出した。湯川の教え子である理論物理学者の武谷三男(1911-2000年)はその座談会で「われわれは記念館とは一体何を作るんだ。神社を建てて、湯川さんというご神体をかざるだけか(笑声)。建物だけで一体何をするんだと。」[13]と語っている。

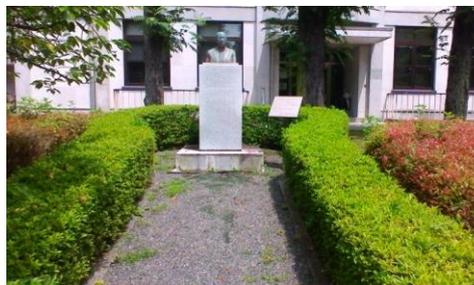

図1　(a)　京都大学基礎物理学研究所湯川記念館前の湯川秀樹胸像遠景。
　　(2018年6月26日　筆者撮影)

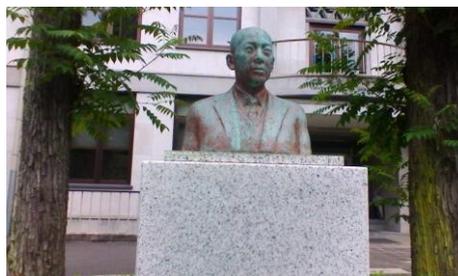

図1　(b)　1981年9月8日設置の湯川財団寄贈湯川秀樹胸像近景(京都大学)。
　　(2018年6月26日　筆者撮影)

基礎物理学研究所・湯川記念館はヨーロッパの量子論発祥の地であるデンマーク・コペン

---

‡　座談会「基礎物理学研究所をめぐって」I.　建設時代(湯川記念館として)(1957年10月13日、京都・清風壮にて)には、小林稔(京都大学理学部教授)、湯川秀樹(基礎物理学研究所所長)、長谷川万吉(京都大学名誉教授)、武谷三男(立教大学理学部教授)、中村誠太郎(東京大学理学部助教授)、高木修二(大阪大学理学部助教授)、吉田思郎(東京大学原子核研究所助教授)が出席している[13]。

ハーゲンのニールス・ボーア研究所[§]のように国際研究拠点となり世界の最先端の研究者があつまってくる理論物理学の＜メッカ＞になっている今日では信じがたい＜神社論議＞である。湯川記念館前の胸像設立は湯川秀樹先生が 1981 年 9 月 8 日に世を去られた 5 年後のことであるから、当然湯川先生は御自身の胸像を見ているはずがないと思っていた。

　だが、湯川先生は生前にしかも若い頃に「湯川胸像」をご覧になっていたのである。京都大学に胸像が建てられるよりも 32 年以上も前に、高知県の小学校においてである。湯川先生の胸像は京都大学のもののみと思っていたが、それよりもはるか前に最初の胸像が筆者のふるさと高知に設置され湯川先生ご夫妻が出席して除幕されていたのである。

　戦後の日本人にとって、日本人としてはじめてノーベル賞を受賞し日本国民におおきな希望と勇気をあたえた湯川先生の名声は今日では想像しがたいほど大きいものであった。湯川先生が米国から帰国後まもなく高知を訪れ大歓迎をうけたこと、先生の胸像がわたしの父の生家(南国市)に隣接する現在の香南市の町につくられたことなど、学校の理科の教師(高知市立昭和中学校、現在の城東中学校)であった父(大久保幹生)はとうぜん職場や報道などで知っていたと思われる。筆者は両親・祖父母から湯川の訪問や銅像について聞かされたのか記憶にはさだかではなく、今となっては世になく訊くこともならず残念である。私の父の自伝にもこのことは記されてない。

　高知に胸像が設立され湯川先生が臨席されたとき筆者は小学校 1 年生である。多くの若者が湯川先生の偉大な存在と名声に影響をうけたように、筆者も湯川先生に憧れて京都大学に進み先生の教えをうけ理論物理学、原子核理論の研究者の道に進むことになったが、小学生の当時には思いもよらないことである。私のふるさと高知に日本で最初に建立された湯川先生の胸像であるので、湯川先生に教わったころをなつかしく思い出しながらすこし調べてみることにした。

## 2. 日本で最初の高知の湯川秀樹先生の胸像

　湯川先生が胸像除幕式出席のおりはじめて訪れた高知市の桂浜は月の名所でも知られる景勝の地である。ここには 1928(昭和 3)年高知県の青年有志によって建てられた幕末の志士・坂本龍馬(1836(天保 6)-1867(慶応 3)年)の高さ 13.5 メートルの巨大な銅像(本山白雲作)がある。雄大な太平洋に向かって立つ銅像は龍馬に広い世界の存在を知らしめ雄志をかきたてたジョン・万次郎(中濱万次郎)[**](1827(文政 10)-1898(明治 31)年)が渡ったアメリカ大陸を見つめているようである。1935 年核力を引き起こす新粒子の存在を予言する中間子論を建設し素粒子物理学を切り拓いた湯川秀樹先生も夫人とともにこの銅像を見上げたことであろう。その 1935 年、海援隊長の龍馬と京都河原町三条の近江屋で運命をともにする

---

[§] 湯川は 1948 年 12 月 10 日のストックホルムでのノーベル賞授賞式のあと、12 月 15 日ニールス・ボーア研究所(当時は理論物理学研究所)を訪問し、17 日まで滞在している。
[**]地元の有志により銅像(西常雄作)が 1968 年、足摺岬先端に太平洋に向いて建てられている。

陸援隊長・中岡慎太郎(1838(天保9)-1867(慶応3)年)の銅像(本山白雲作)が出身地元の青年が中心になって室戸岬の先端に建てられ、龍馬とともに太平洋を見つめている。この地、室戸岬は湯川秀樹が「万能的な天才」††[14]と語る真言宗の開祖弘法大師・空海(774(宝亀5)-835(承和2)年)が御厨人窟(みくろど)で修行し神明窟(しんめいくつ)で悟りを開いたといわれ、空と海が地平線に一体・無に見える。龍馬の生地は高知市上町、中岡慎太郎は室戸に近い安芸・北川村である。坂本龍馬や中岡慎太郎とともに近代日本の建設に貢献した三菱の祖、岩崎弥太郎(1835(天保5)- 1885(明治18)年)も安芸の出身であり、銅像が1985(昭和60)年安芸市に生誕150年を記念し有志により建てられている。岩崎弥太郎の銅像は長崎県・高島にもあり、グラバー商会で幕末・維新に活躍し炭坑開発でも貢献した中岡慎太郎と生年が同じのトーマス・グラバー(Thomas Glover, 1838(天保9)-1911(明治44)年)の故国英国のある西方に向いて立っている。グラバーは英国人の物理学者ラザフォード（E. Rutherford, 1871(明治4)-1937(昭和12)年)が39歳で原子核を発見[17]した1911(明治44)年の暮れに日本で死去した。

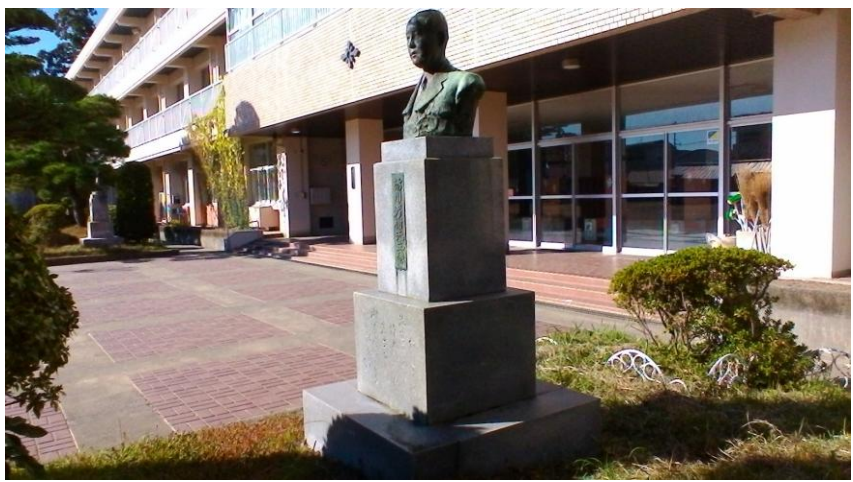

図2　1954年3月建立された高知県夜須小学校の湯川秀樹胸像遠景。
（2018年11月4日　筆者撮影）

ラザフォードが発見した原子核のなかで陽子と中性子にはたらく謎の力である核力の成因をラザフォードが66歳で世を去る2年前の1935年に28歳で中間子論の論文を発表[3]し解明した湯川秀樹先生の日本で最初に建てられた胸像は安芸市西方の農漁村の町、夜須町にある(図2、図3)。湯川秀樹は歌人でもあり多くの和歌を詠んでいる[18]が、日本最初の仮名日記文学である『土佐日記』[19]の作者、歌人・紀貫之(866(貞観8)(あるいは872(貞

---

††湯川は弘法大師・空海(四国八十八箇所霊場を開創)について著書『天才の世界』[14]のなかで「ひじょうに万能的な天才の特色がある。彼は自分の思想体系を構築し・・・その体系のなかに他の思想はみなはいってくるわけです。日本では珍しい体系的思想家でもあった。」(p.28)と述べ、3冊の『世界の天才』[14]、『続世界の天才』[15]、『続々世界の天才』[16]の天才シリーズで最初に弘法大師・空海をとりあげている。

観 14)-945（天慶 8)年、諸説あり）が土佐での任務ののち京へ帰る途中に立ち寄った奈半利は夜須町の少し東方にあり、紀貫之は 935（承平 5)年 1 月 9 日に夜須町の風光明媚な住吉海岸、手結の海を通過しながら眺め、

　　　　見渡せば松の末ごとに住む鶴は千代のどちとぞ思ふべらなる

と歌に詠んでいる[19]。47 歳の湯川秀樹先生と夫人湯川澄子さん(1910-2006 年)[20]はこの夜須町を日本ではじめての湯川秀樹胸像の除幕のために訪れた。5 年間の米国生活から帰国されてまもなく 1954 年 3 月の春である。

　湯川先生の胸像建設は、当時、小学生、中学生、高校生、住民であった夜須町および近隣市町村の人々や高知県内の戦前生まれの一定の年齢以上の人々にはもとより広く知られていることである。古い新聞などをめくってみると湯川先生夫妻は 1954 年春、私の住む小高坂からすぐ近くの高知市上町の城西館という旅館に泊まっていることがわかった。城西館は坂本龍馬が泳いだ鏡川に近く、また生誕地である同じ高知市上町内にある。天皇家、皇室が定宿として利用している旅館としても知られ、地元の人にも親しみのある老舗旅館ある。湯川先生が高い待遇で宿泊したことがわかる。

　さらに古い年鑑[21]や新聞などから調べをすすめると、日本人として初めてノーベル賞を受賞し国民にとって英雄であった湯川先生が高知市内の代表的な学校ではなく、郡部の小学校を訪れていたことがわかった。高知市から室戸方面に 20km ほど行った小さな農漁村にある夜須小学校である。当時はこんにちほど高速道路網も発達してなく交通の便はよくない。現在の香南市夜須町夜須小学校で、当時の高知県香美郡夜須町の夜須小学校である。2018 年 9 月初め頃までに湯川訪問の概要がわかった。私の祖父（小味政吉）は香南市（赤岡）の出身で当地のことは母（大久保政子）からもよく聞いた記憶があり、湯川先生がその地を訪ね講演もしたと知りいっそう身近に感じられ大変うれしく次の拙歌を詠んだ。歌人吉川宏志[22]主宰の京都の短歌誌『塔』に投稿し 3 か月後の 2018 年 12 月号に歌人小林幸子[23]の選歌で掲載された[24]。

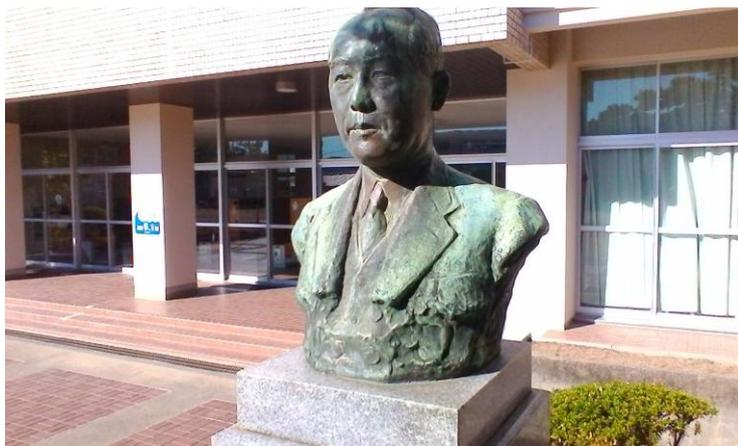

図 3　夜須小学校にある日本で最初の湯川秀樹先生胸像。
（2018 年 11 月 4 日　筆者撮影）

胸像の除幕式にと京をたち夫妻は訪ねし木造校舎

敗戦の落胆日本に希望あり偉人は名もなき小学訪ぬ

湯川と交わりのあった素粒子物理学者の小沼通二さんに伝えると、夜須小のことは未知でぜひ訪問したいとのことで、後日、湯川胸像写真も送付した。湯川の高知訪問の詳細は 2018 年 9 月中に把握できた。できるだけ早期に写真ではなく、夜須町を訪れ実物の湯川胸像を見てみたいと思ったが、夜須町は遠く 10 月中の訪問もかなわなかった。10 月初旬に、筆者が 50 年近くかかわってきた原子核のクラスター構造研究に関して、池田清美博士(1934-)の発見した「Threshold rule」（閾値則）[25]の 50 周年を記念する京都大学基礎物理学研究所の研究集会「Threshold rule 50」(10 月 3 日—5 日)が開催されることになっていた。原子核のクラスター構造研究は湯川中間子論と核力の研究を踏まえ、日本の研究が世界をけん引した研究分野であり、湯川先生も高く評価され、また、湯川生誕 100 年をひかえたアインシュタインの相対論 100 周年記念の 2005 年にひらかれた理論物理学を中心とした分野横断型の基礎物理学研究所の研究シンポジウム「学問の系譜－アインシュタインから湯川・朝永へー」[26]‡‡でも原子核のクラスター研究が日本で発展した独創的研究分野として取り上げられた。研究集会「Threshold rule 50」は日本と世界のクラスター研究をけん引してきた池田清美先生の業績を記念するとともに今後の発展を展望する研究集会であったので、おおくの古参研究者・若手とともに筆者もこれに参加、講演し[28]、また引き続いて 21 年ぶりのブルガリア訪問、はじめてのルーマニア訪問[29]など、9 月-10 月は銅像を見に行く機会を得られなかった。

2018 年 11 月 4 日、筆者は湯川秀樹胸像を見るために夜須小学校を訪れることができた。湯川が 65 年前に訪れたときの木造の校舎は、白い鉄筋コンクリートの立派な建物にかわっていた(図 2)。玄関右わきの松の木のもとにひっそりたたずむ湯川先生の銅像に対面し、京都大学で教えを受けた若いころを思い出し感慨深いものがあった。このとき詠んだ和歌を短歌誌『塔』に投稿し 3 か月後の 2019 年 2 月号[30]に歌人前田康子[31]の選歌で掲載された。

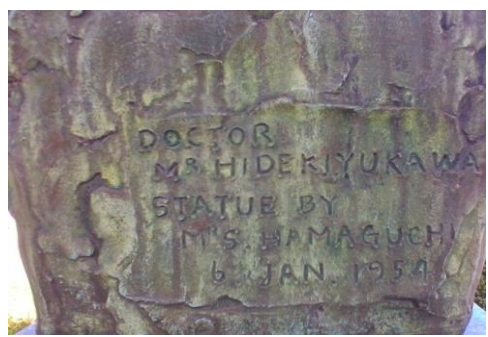

図 4　湯川秀樹胸像の裏面（香南市夜須町夜須小学校）。(2018 年 11 月 4 日　筆者撮影)

---

‡‡ 原子核物理、宇宙線、基礎物理学、宇宙物理学、物性物理学、生物物理学、素粒子論、自然の累層構造、と物理学全体にまたがるシンポジウムで、南部陽一郎(1921-2015 年)は「基礎物理学の系譜：基礎物理学−過去と未来」[27]と題し、また京都大学基礎物理学研究所所長の九後太一は「場の理論の発展と日本」[27]と、それぞれ壮大な演題で講演した。

松陰にひっそり鎮座の銅像に帽子をとりて頭(かうべ)を垂れり

　このときの感激は筆者が 1983 年コペンハーゲンのニールス・ボーア研究所の地下実験室前の廊下に展示されている写真（先頭は紀元前、原子論のデモクリスト(Democritus, BC460 年頃－BC370 年頃)で、コペルニクス(N. Copernicus, 1473-1543 年)、ガリレオ(Galileo Galilei, 1564-1642 年)、ニュートン(I. Newton, 1642-1727 年)、マクスウェル(J. Maxwell, 1831-1879 年)、アインシュタイン(A. Einstein, 1879-1955 年)、など現代までの世界の偉大な科学者 55 人の写真で、最後は湯川秀樹、湯川以外はすべて西洋人）のなか、最後の 55 番目に、唯一東洋人として、湯川先生を発見したときの感激に劣らぬものであった[§§]。夜須小学校の湯川胸像は京都大学基礎物理学研究所の銅像(図 1)に似ている感じもするが、65 年の風雪を感じさせるものであった。胸像の裏面(図 4)には

```
DOCTOR     MR    HIDEKI    YUKAWA
STATUE    BY   MR    S.HAMAGUCHI
6    JAN   1954
```

とある。1954 年 1 月 6 日彫刻家浜口青果（本名、重蔵、1895-1979 年）[34]によって制作されたことがわかる。浜口青果は高知城公園にある板垣退助の銅像制作でも知られている。

## 3. なぜ湯川秀樹胸像が夜須小学校に建立されたか

　なぜ夜須小学校に胸像が設立されることになったのか。その趣旨は胸像裏面の銅板に刻まれた「設立趣意書」（図 5)に詳しく書かれている。胸像裏面の設立趣意書の刻文字は歳月をへて判読しづらいところもあるが、図 6 のようである。カタカナを平仮名にし、文と文の間に空白を入れて読みやすくすると次のようである（原文は縦書き）。

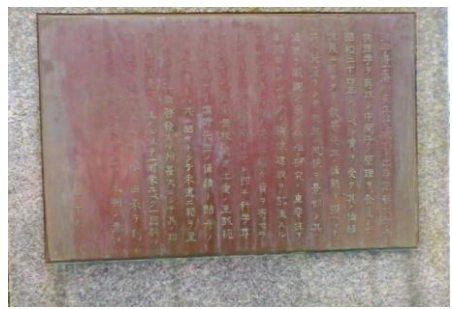

図 5　湯川秀樹胸像裏面の銅板設立趣意書。
　　（2018 年 11 月 4 日　筆者撮影）

「理学博士湯川先生は東京都の出身
京都大学に物理学を専攻し中間子の原理を発見して昭和二十四年ノーベル賞を受く　其の偉績世界に赫赫たり　我等先生の偉勲を讃へると共に児童をして先生の風貌を景仰し其の偉業を感謝し深く科学研究の重要性を感銘せしめんがため胸像建設を計画するや町出身縣内外の先輩資を寄せらる　町民又協力し像なる　是れ即ち科学尊重の精神児童愛母校愛郷

---

[§§]　この時の印象は「徳島科学史研究会創立 30 周年」記念講演、「原子核発見 100 年をむかえて」（「徳島科学史雑誌」[32]）および素粒子論研究電子版[33]に収録されている。

土愛の至誠純情の結晶なり　嗚呼先生の偉業と諸彦の熱誠芳情とは共に昭々として永遠に範を垂る　後進を奨励啓発する所甚大にして其の功徳俱に無量なり　此の心以て一町栄ゆべく一国興るべし　感激に堪へず仍て健像の由来を刻して之を不朽に傳へ深甚の感謝の意を表す」

　格調高い漢文調である。胸像は子供たちのために建てられた。設立趣意書は次のように刻んでいる。

「児童をして先生の風貌を景仰し其の偉業を感謝し深く科学研究の重要性を感銘せしめんがため胸像建設を計画する」。

　事実、銅像の台座一番下の表側（図7）には設立主体のPTAから夜須小学校の子どもたちにあてたメッセージが子供たちのわかる文体で刻まれている。風雪をへて判読しにくいので書き下すと図8のようである。この胸像が子どもたちのために作られたことがよくわかる。

　この胸像は特定の個人・団

図6　夜須小学校湯川秀樹胸像裏面に記された胸像設立趣意書の解読文。

理学博士湯川先生ハ東京都ノ出身京都大学ニ物理学ヲ専攻シ中間子ノ原理ヲ発見シテ昭和二十四年ノーベル賞ヲ受ク其ノ偉績世界ニ赫々タリ我等先生ノ偉勲ヲ讃ヘルト共ニ児童ヲシテ先生ノ風貌ヲ景仰シ其ノ偉業ヲ感謝シ深ク科学研究ノ重要性ヲ感銘セシメンガタメ胸像建設ヲ計画スルヤ町出身県内外ノ先輩資ヲ寄セラル町民又協力シ像ナル是レ即チ科学専重ノ精神児童愛母校愛郷土愛ノ至誠純情ノ結晶ナリ嗚呼先生ノ偉業ト諸彦ノ熱誠芳情トハ共ニ昭々トシテ永遠ニ範ヲ垂ル後進ヲ奨励啓発スル所甚大ニシテ其ノ功徳俱ニ無量ナリ此ノ心以テ一町栄ユベク一国興ルベシ感激ニ堪ヘズ仍テ健像ノ由来ヲ刻シテ之ヲ不朽ニ傳ヘ深甚ノ感謝ノ意ヲ表ス

PTA

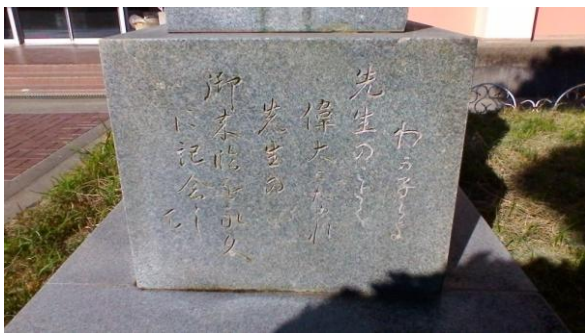

図7　湯川秀樹胸像台座一番下、子どもらへのメッセージ碑の写真。(2018年11月4日　筆者撮影　香南市夜須町夜須小学校)

わが子らよ
　先生のごとく
　　偉大になれ
先生の
　御来臨を永久
　　に記念し

図8　湯川秀樹胸像台座の一番下（図7）に刻まれた子供らへのメッセージの文面（筆者解読）。

体の大型寄付金ではなく、自発的な住民の科学研究・科学教育振興の願いで設立されてい

る。夜須小学校 PTA の呼びかけに応じて胸像建設には多くの町民が募金している。また、胸像設立には夜須町出身の県内、県外の多くの人々の自発的寄付があったことを、設立趣意書は次のように記している。

**「我等先生の偉勲を讃えると共に児童をして先生の風貌を景仰し其の偉業を感謝し深く科学研究の重要性を感銘せしめんがため胸像建設を計画するや町出身縣内外の先輩資を寄せらる　町民又協力し像なる」**。

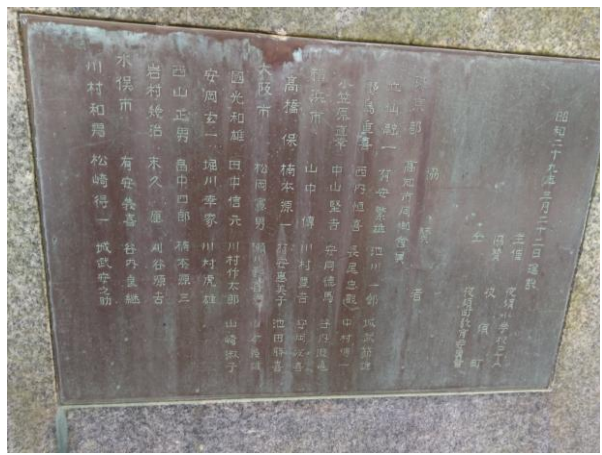

図 9　湯川秀樹胸像の裏面に記された設立協賛者。
（2018 年 11 月 4 日　筆者撮影）

胸像の裏面(図 9)にはその寄付者の氏名が刻まれている。協賛者をみると夜須町以外の在住者の氏名が記され、東京都、横浜市、大阪市、水俣市にわたっている。風雪で劣化し判読しがたいものもあるが、図 10 のように読めた。『夜須町史』[35]をみると協賛者の親族・子孫が現在も夜須町に住んでいるようである。

```
昭和二十九年三月二十二日　建設
主催　夜須小学校　ＰＴＡ
協賛　夜須町

協賛者
東京都　高知市同郷会員
　立仙融一　有安繁雄　池川一郎　城武節雄
野島重喜　西内恒喜　長尾忠観　中村傳一
　小笠原直幸　中山堅吉　安岡徳馬　谷内遊亀
横浜市　山中傳　川村豊吉　安岡友喜
高橋保　楠本源一　有安恵美子　池田勝吉
大阪市　松岡憲男　瀬川壽喜治　山本義雄
　国光和雄　田中信元　川村作太郎　山崎淑子
　安岡玄一　堀川幸家　川村虎雄
　西山正男　畠中四郎　楠本源三
　岩村幾治　末久　愿　刈谷源吉
水俣市　有安義喜　谷内良継
　川村智男　松崎得一　城武安之助

全　夜須町教育委員会
```

図 10　湯川秀樹胸像設立協賛者名（銅板図 9)。（筆者解読）

浜口青果に胸像制作を依頼したのは夜須小学校 PTA 川村春吉会長であり、青果が夜須町に一時居住した縁で依頼している[36]。この川村春吉会長こそ湯川胸像の設置を思い立った発起人である。当時の新聞[36]によると「かねてから科学教育の振興を熱心に唱えてい

た香美郡夜須小学校 PTA 川村春吉会長は子供たちがあやかるようにノーベル賞を受けた湯川博士の胸像をたてることを思い立ち昨年同地に一時居住していた彫刻家浜口青果さんに制作を依頼」したとある。胸像制作は 1953 年 10 月には始まっている[37]。湯川が日本に帰国するのは 1953 年 7 月であるので、帰国まもなく胸像制作が開始されていたことになる。1953 年 12 月 6 日（日曜日）の高知新聞朝刊「町から村から」欄は「夜須町小学校 PTA 委員会が湯川秀樹博士胸像について協議」[38]と報じている。

夜須町史編集委員会が纏めた『夜須町史』（上巻）[35]第 4 章 p.645 の「町立夜須小学校沿革」に次のように湯川秀樹の訪問の記述がある。

「昭和二十四年　近森亘校長着任　ＰＴＡ発足、11 月図書館発足／　昭和二十九年　ノーベル賞受賞者湯川秀樹博士胸像設立除幕式、浜口青果制作、経費 10 万円」。
町史にある昭和 24（1949）年に発足した PTA というのが重要である。湯川秀樹胸像を設置したのはこの夜須小学校 PTA が主体となった運動である。PTA は戦後日本を占領支配したアメリカ占領軍が四国でもその運動を奨励したものである。戦後の教育の民主化運動のなかで夜須小学校に PTA が設立された。

夜須町にかぎらず高知には偉大な人物を尊敬し子どもたちや後世に伝えるため有志で銅像をつくるなどの熱情が感じられる。隣接する夜須中学校の正門横には教育者・教育学者でアメリカ・プラグマティズムのデューイの学説による「西山式」とよばれる学習指導法を確立し夜須小学校でも教えたことのある西山庸平[39]（高知県南国市田村出身、1872 – 1939 年）をたたえる浜口青果作の胸像が建てられている。土佐の民権運動思想家・植木枝盛（1857-1892 年）の「自由は土佐の山間から」という自由民権思想・運動や、青年有志による坂本龍馬の銅像設立など、土佐の民衆運動が底流にあるようだ。

## 4．湯川胸像除幕式(1954 年 3 月 22 日)と湯川秀樹夫妻

3 月 22 日夜須小学校での湯川博士胸像除幕の写真が図 11 に示されている。当時の新聞記事によると[37]「この銅の胸像の設立は同校 PTA の発案により昨年十月から同町出身の彫刻家浜口青果氏が制作にとりかかり戦時中からあった楠公の胸像に変ってこのほど同校の正面玄関の右側に完成したもの」とある。さらに、「この日式場へは朝早くから同小学校児童七百名、同中学校生徒二百名をはじめ県教委会指導課の大倉上田両氏、橋詰延寿氏、鍋島市公民課長ら来賓、町民約三百名が参列、PTA 会長川村春吉氏の式辞にひきつづき、同校六年生高橋南海男君と春樹英子さんの二人によって幕がとり除かれた、つづいて川村会長から感謝状が制作者浜口氏の代理長男重敬君（土佐高二年）らに贈られ」[37]たとある。

湯川胸像が楠公にとってかわって、また、玄関左側の二宮尊徳（二宮金次郎）と対置して設置され、その偉人としての格段の扱いぶりがわかる。湯川像にとってかわられた「楠公」（楠正成）はいうまでもなく湊川神社（神戸市）に神として祭られ、日柳燕石（1817（文化 14）-1868（慶応 4）年）作『楠公を詠ず』「日東に　聖人有り／其の名を　楠公と曰う／誤って　干戈の世に生まれ／剣を　提げて　英雄と作る」[40]や徳川斉昭（1800（寛政 12）-1860 年（万延元）

年)作『大楠公』「豹は死して皮を留む/豈偶然ならんや/湊川の遺跡　水天に連なる/人生限り有り　名は尽くる無し/楠氏の精忠　万古に伝う」[41]など漢詩や詩吟剣舞で現在でも広く語り継がれている。二宮尊徳も報徳二宮神社(小田原市、日光市)に神として祭られてい

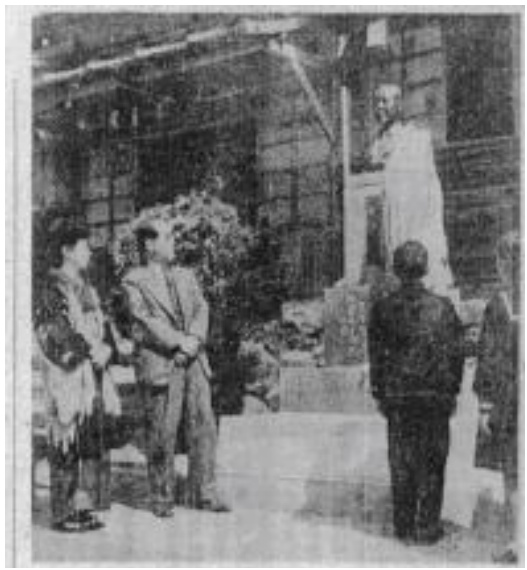

図 11　夜須小学校の湯川秀樹胸像除幕式に出席の湯川秀樹、湯川澄子夫妻(左)。除幕したのは写真右側の夜須小学校六年生高橋南海男君と春樹英子さん[37]。校舎は木造である。

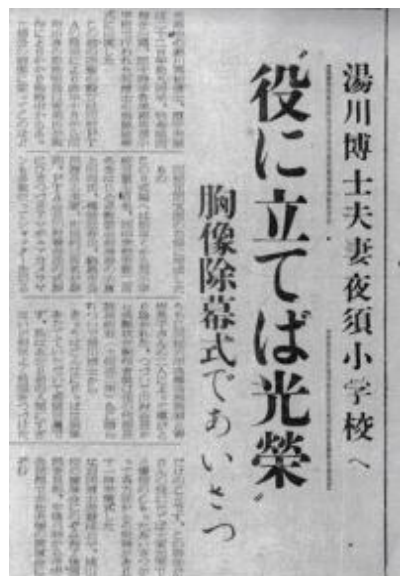

図 12　湯川胸像除幕式を報ずる高知新聞 1954 年 3 月 23 日（水曜日）夕刊)[37]。

る。戦後、楠公や二宮尊徳は教科書から消えていったが、戦前の教育での位置づけを考えると一般の人々の湯川秀樹先生にたいする尊敬・尊崇の念がきわめて高いことがうかがい知れる。

　湯川秀樹は夜須小学校での除幕式でのあいさつで、「きょうはこんなにりっぱな胸像をたてていただいて感慨無量です、私はあたり前の人間にすぎないが根気よく勉強をつづけただけのことです、この胸像がみなさんの役にたてば大変光栄です」[37]と述べている。図 12 のように新聞は大きく「役にたてば光榮」と見出しで伝えている。

　二宮尊徳の銅像とならんで胸像が立つことについて湯川先生は 1954 年 3 月 21 日夕刻高知駅に到着時の記者会見で記者の質問に答えて次のように語っている。「私は皆がいうほど偉人でもなければりっぱな人間でもない。二宮尊徳像にかわって私の胸像建立

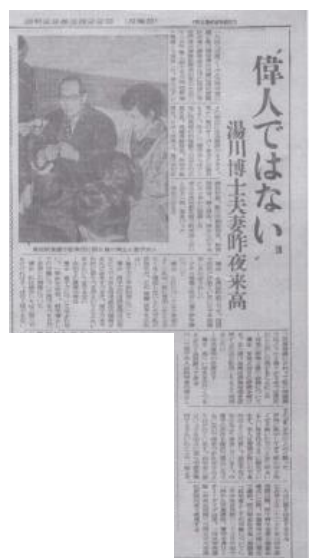

図 13　高知到着時の湯川秀樹の記者会見を報ずる高知新聞(1954年 3 月 22 日朝刊)[42]。

はどうかと思うが日本では最初のことだし出席することにした」[42]。この会見を報じた新聞は「"偉人ではない"」という大きな見出しをつけて紹介している(図13)。はじめて直接対面し取材するたぶん雲の上のひとである湯川先生ご本人から出たことばに、新聞記者はさぞかし驚いたに違いない。湯川先生の人柄を知る研究者や関係者なら、湯川先生がひかえめな性格であることは自伝『旅人』[43]などで知られていることであるが、この新聞見出しをみると新聞記者ならずとも、一般の県民・市民にとってもたいへんな驚きであったことだろう。

　湯川は3月23日の高知市での学童向けの講演で「偉人」について「何でも出来る人は何でもたいしたことはない人だと私は思います。偉い人とはどういう人をいうのか疑問ですが、いわゆる偉い人になるのが問題でないと思います。人にはそれぞれの特徴があります。その特徴、自分にはどんな可能性があるのかを知って伸ばし世に役立つことが問題であります。」[44]と語っている。浮き世を超越したようにみえる素粒子という理論物理学研究をおこないながら、「世に役立つこと」いうのが根底の信念としてあり、夜須小学校の胸像設立除幕式への出席に賛同したのではないだろうか。

## 5．1954年ごろの湯川秀樹の名声

　湯川秀樹はいまや国語の教科書にも出てくる歴史上の人物で直接に接した人は高齢化し少なくなりつつある。京都大学で湯川先生の講義を直接受けた世代は私の1学年後までである。敗戦直後日本人として初めて1948年ノーベル賞を受賞した湯川先生について、私の2006年に記された「湯川先生の思い出」[45]には次のようにしるされている。「ストックホルムからのノーベル賞の朗報は敗戦で焦土と化し、食糧難など日々の生存すら厳しい状況におかれ生活困窮と精神的落胆でうちひしがれていた日本国民におおきな希望と勇気をあたえました。この国民にあたえた勇気と自信はその後の日本の戦後復興、経済的発展の精神的原動力になったとも言われています。理論物理学の素粒子・原子核研究は国民の大多数にとって日々の生活に直接関係のない学問分野です。このような分野での研究がこのように国民におおきな生きる力をあたえるとは、基礎科学のもつはかりしれない文化的な力の大きさを感じさせられます。湯川先生は一躍国民的英雄となりました。幼児であった私にはこのときのことは知るよしもありませんが、地方で生まれた私の運命にも関係してくるとはその影響の大きさを思わずにはいられません。」

　湯川秀樹の偉大さが1954年当時国民にどのように受けとめられていたか、いくつかの側面から見てみよう。(1)大阪市出身の文豪・川端康成(1899(明治32)-1972(昭和47)年)、1968年ノーベル文学賞受賞)の小説の記述と、(2)1953年理論物理学国際会議で日本を訪問したオランダ人（のち米国籍）素粒子物理学者アブラハム・パイス（Abraham Pais、1918-2000年）が湯川秀樹を「天皇についで影響力のある」と記した本、および(3)当時の学校の理科教科書、をとりあげてみる。

### （１） 川端康成の新聞連載小説の湯川秀樹－「湯川博士は敗戦日本の栄光で、希望」

第１は一般市民の目線である。川端康成は小説『舞姫』（図14）を朝日新聞に1950(昭和25)年12月12日から翌1951(昭和26)年3月31日まで109回連載で発表し、その中に湯川秀樹を実名で登場させた。湯川が米国にプリンストン高等研究所客員教授として出発したのは1948年9月1日、9月3日に米国サンフランシスコに到着。東海岸プリンストンで研究生活にはいり、翌1949年8月にはコロンビア大学の客員教授となりニューヨークに移る。同年11月3日ノーベル物理学賞の受賞が決定し、12月10日ストックホルムでノーベル賞受賞式に出席、12月29日にはニューヨークに戻る。翌1950年8月湯川は受賞後初めて一時帰国し、国民の大きな歓迎を受けている。9月1日朝10時羽田発パン・アメリカン「ゴールデンゲート号」で米国コロンビア大学へ帰任[46]。川端康成が『舞姫』の連載を開始するのは、1950年12月であるから川端が小説連載期間中、湯川は日本にはいなく、米国で研究生活を送っている。小説にでてくるのはこの1950年8月の湯川の帰国の様子を記述したものである。

角川文庫による『舞姫』[47]の湯川秀樹にかんする記述は２か所で次のようである。第１箇所目は小説の最初の方（p.39－41）。

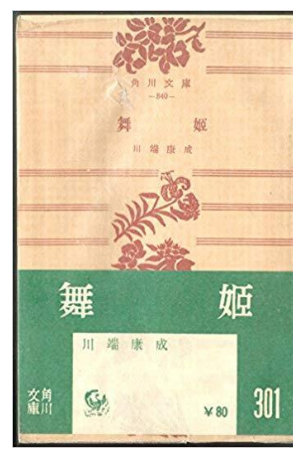

図14　ノーベル賞湯川秀樹の帰朝を描写した川端康成の新聞連載小説『舞姫』[47]。

　「どうだ、落ちつくだらう。宿屋になる前は、鉄成金の家で、ここは茶屋だった。あの、ノオベル賞の湯川博士も、この部屋に泊つてゐたんだ。アメリカから飛行機で着いた時にも、アメリカへ飛行機で立つ時もね・・・・・・。水泳の古橋選手らも、アメリカの行き帰りに、ここで合宿したんだ。」

　「お母さんが、よくいらつしゃるうちぢやないんですか。」

　と、高男は言つた。

　湯川博士や古橋選手は、敗戦日本の栄光で、希望で、その人気ものが、アメリカの行き帰り、泊まつたといふ部屋に、通されたのだから、若い学生は心ときめかせるだらうと、矢木は思つたが、高男はそれほどに感じないらしかつた。

　矢木はつけ加へた。

　「ここへ来る手前にね、広い部屋があつたらう。あの二間を打ち抜いて、湯川博士の面会室にあてたんだ。いろんな人の押しかけてくるのを、なるべくこの居間には、通さんやうにしたんだが、新聞社の写真班が、どこからともなく、庭にしのびこんで来て、変つた姿をねらふから、湯川さんは、ほつとくつろぐ間がありやしない。写真班を入れないやうに、ここの女中が二人、庭の両端で、夜も張り番に立つてゐて、蚊にさされて困つたさうだ。夏だったからね。」

　　矢木は庭に目を向けた。
　　大名竹、布袋竹、寒竹、四方竹など、竹ばかり植ゑこんだ庭で、隅に稲荷の赤い鳥居が見えた。この部屋も竹の間と言ひ、すす竹の天井であつた。
　　「湯川博士が着いた時、宿の奥さんは病気だつたんが、久しぶりで日本にお帰りになつたんだから、いい香をたいて、朝顔も咲いてゐると、寝ながら気をくばつて、庭の木に、せみも鳴いてくれればいい。」
　　「はあ・・・・・・。」
　　「せみも鳴いてくれればいいとはおもしろいね。」
　　「はあ。」
　　しかし、高男は同じ話を、前に母から聞いてゐた。父は母の受け売りをしてゐるらしいので、息子はおもしろい顔がしにくかつた。
　　部屋を見まはしながら、
　　「いいうちですね。お母さんは、今でも、よく来るんでせう。ぜいたくなんだな。」
　　父は吉野丸太のしぼり手の床柱を背に、ゆつたりと座つて、うなづきはしたが、
　　「せみは鳴いてくれたらしいんだね。東京の、宿に来て先づ、なつかしむ、せみの声する、庭の木立を。その時の湯川博士の歌だ。湯川さんには、かねて歌のたしなみがある。」

湯川秀樹にかんする記述の2箇所目(p.232)は小説の終わりの方で次のようである。
　　つきあたりの庭に、植木屋が、枯松葉を敷いてゐた。
　　そこを右にまがり、また左に折れて、湯川博士が泊つた、竹の間の裏から、庭に出て、
　　「矢木が来た時、そのお部屋でしたつて・・・・・・？」（下線は筆者）

川端康成が湯川秀樹を「日本の栄光」と記述するのも、当時の日本の国民の受けとめをあらわしている。小説に出てくる旅館幸田屋で帰国時に湯川の読んだ和歌は川端が引用した

**東京の宿にきて先ずなつかしむ蝉の声する庭の木立かな**

のほかに

**東京の宿の障子に小窓あり竹の葉末のゆれて光れる**

がある[48]。

湯川が2度目の一時帰国をするのは1952年7月20日であり、この時には『舞姫』は単行本(朝日新聞社、1951年7月)として書籍化されている。『舞姫』は新藤兼人脚本、成瀬巳喜男監督・演出により山村聡、高峰三枝子出演で映画化され(東宝)、1951(昭和26)年8月に封切公開されている。また、文庫版『舞姫』が角川文庫で発売されたのは1954(昭和29)年7月、新潮文庫は1954年11月である。

　（２）　物理学者パイスの見た湯川秀樹－「湯川は日本で天皇についで有名」
　第2はプロの研究者が湯川をどう見ていたかである。外国人研究者の見方が率直でわかりやすい。理論物理学者・素粒子物理学者であるパイスは著書『物理学者たちの20世紀　ボーア、アインシュタイン、オッペンハイマーの思い出』[49](図15)のなかで、湯川に対す

る日本国民の受けとめ方について、1953年秋日本で開かれた湯川の随筆『しばしの幸』[1][2]にもでてくる理論物理学国際会議にかんする記述のなかで言及している。パイスの記述の理解をたすけるため、この前後の湯川の動向を簡単に紹介する。

　湯川が5年間の米国生活から東京羽田に帰国するのは1953年7月16日(木曜)午後6時半である[50]。ヨーロッパ経由の長旅で、羽田空港には教え子の物理学者である小林稔京都大学教授(1908-2001年)、坂田昌一名古屋大学教授(1911-1970年)らが出迎えている[50]。湯川の帰国からまもなく1953年9月11日-24日、日本で最初の国際会議である理論物理学国際会議が京都と東京、日光の会場で開催された。湯川はこの国際会議の会長を務めている。湯川の帰国は敗戦後初めての国際会議である理論物理学国際会議を成功させることが大きな仕事であった。湯川は1953年7月1日付けで京都大学教授に復帰し、8月1日には日本で初めての全国共同利用研究所として京都大学に付置研究所として設置された湯川記念館・基礎物理学研究所の所長に任命されている。9月1日には京都大学理学部教授から基礎物理学研究所教授に配置換えになり、理学部物理教室は併任教授となる。湯川教授室はその後も理学部物理学科にそのままあった。国際会議は朝日新聞の社説でもとりあげられ、日本全体にとっても大きな国際復帰の出来事であった。海外から55名が参加し、そのなかにはファインマン(R. P. Feynman, 1918-1988年)、ウイグナー(E. P. Wigner, 1902-1995年)、マイヤー(M. G. Mayer, 1906-1972年)などノーベル賞級の学者がたくさん含まれていた(ノーベル賞受賞者は後も含めると17名)。パイスは9月13日に東京に到着後、皇居を訪問、首相官邸も訪問した。外務大臣と文部大臣によるパーティにも出席したことを記し、国を挙げて湯川の国際会議をバックアップしたことがわかる。パイスは急に来られなくなった、理論物理学者で「原爆の父」とよばれアメリカの原爆開発を指導したオッペンハイマー(R. Oppenheimer, 1904-1967年)の代わりに出席することになったことを著書[49]で次のように述べている「オッペンハイマーが直前になって学会出席の予定を取りやめた・・・開会式の議長もできなくなった・・私(パイス)が代わりに選ばれた」。

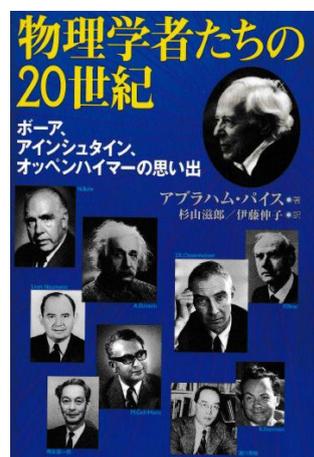

図15　「湯川は日本で天皇についで有名」と述べたパイスの本[49]。

　パイスは京都での理論物理学国際会議のあい間に、京都の店に美術品を買いに行くにあたって、湯川に紹介状を書いてもらって、店を訪ねたときのことを次のように記している。

> 「私は湯川に、美術品を買えそうな場所を尋ねた。すると彼は、京都の新門前通りにある今井氏の店を訪れるようにと言い、紹介状も書いてくれた。次の日、私は学会をさぼってその店に行った。店に入ると、店主が出てきた。年配の威厳ある人物で、私が紹介状を渡すと、大いに驚いた様子だった。湯川は日本で天皇についで有名な人物だったことに注意していただきたい。私達は、腰をおろしお茶を飲んだ。」[49]（下線は筆者）

美術店を訪れたのは 1953 年 9 月 21 日(月曜日)である。理論物理学会議の京都での会議は 9 月 18 日(金曜日)に始まり、パイスは 9 月 19 日午前研究発表をし、夜ファインマンと祇園の芸者ハウスを訪れ、9 月 20 日(日曜日)奈良へ遠足、東大寺を見物した。9 月 21 日にパイスは古美術店を訪問したのである。

### (3) 1954 年ごろの学校理科教科書と湯川秀樹

第 3 に湯川秀樹が当時の学校の生徒・学童にどう紹介されていたかを教科書を通じて見てみる。1954 年の教科書に湯川秀樹は監修者として載っており、また、中間子論研究が教科書でも紹介され、中学生にも広く知られていたことがわかる。学校図書株式会社『理科 中学校 1 年 上』、『理科 中学校 1 年 下』、『理科 中学校 2 年 上』、『理科 中学校 2 年 下』、『理科中学校 3 年 上』、『理科 中学校 3 年 下』は湯川秀樹監修であることが確認できた。

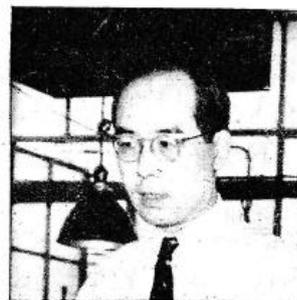

図 16 湯川秀樹の中間子を説明する 1954 の教科書『理科 中学校 3 年下』[51]。(注：湯川の生年は 1907 年 1 月 23 日)。

1955 年には学校図書発行の教科書『小学校理科 4 年上・下』、『小学校理科 5 年上・下』、『小学校理科 6 年上・下』の著者となっている。学校図書株式会社によると小学校 1 年から 3 年の教科書についても同様とのことである。1956 年では英徳社発行の教科書『新制高校生の物理』が湯川秀樹監修である。1957 年には学校図書発行の教科書『しょうがっこうりか 1 ねん』の監修、『小学校りか 2 年』、『小学校理科 3 年』、『小学校理科 4 年上・下』、『小学校理科 5 年上・下』、『小学校理科 6 年上・下』の著者である。

学校図書株式会社『理科 中学校 3 年 下』(図 16)[51]をみると、湯川秀樹の写真、中間子や宇宙線の話が載っており、かなり最先端の話が出ている。関連部分をぬきがきすると、以下のようである。

> 「中間子の発見 電子と,原子核をつくる陽子,中性子の三つが素粒子であって,もうこれ以上に素粒子はないと一時は考えられていた。しかし,原子核の中で陽子と中性子がかたくまとまっている力は,それまでの力では説明できないなぞの力であった。その力を説明するために湯川秀樹は,その後に宇宙線とよばれるようになった新しい粒子が存在すると考えた。そして陽子と中性子の間にはたらく力は,その中間子がなかだちになっているとして説明した。それは 1935 年のことであったが,それから 2 年ばかり後に,世界のあちこちで宇宙線を研究している人々が,中間子に相当する粒子を実際に発見した。それで素

粒子の仲間が一つふえることになった。

　このようにして,自然のしくみを,きわめて少数の素粒子で説明しようという研究は,予想に反して,だんだん考えなくてはならぬ素粒子の数が多くなってしまった。中間子のような性質をもつ粒子も一種類ではないことがわかり,いったん簡単に説明されそうになった物質の根本も,またかえって複雑なものとなってきた。しかし,おそらくは,それは見かけのことであって,新しいものの見方ができれば,この雑多な素粒子も簡単なものに帰着されるであろうというのが,科学者の信念である。」（下線は筆者）

「**宇宙線**　宇宙のどこからくるのかわからないけれども、ラジウムやウラニウムの放射線と違って、宇宙線という放射線が、昼夜の区別なく上空から地上に向けふりそそがれている。・・・・　地球の大気の上層に達すると、酸素や窒素の原子核と作用して、複雑な現象を起こす。

そして,湯川秀樹の予言した中間子や,その他の数多くの新しい粒子をつくり出し,それらがいっしょになって,地上へふってくるのである。」（下線は筆者）

湯川秀樹が学校生徒にも教科書などを通じて知られていたことがわかる。

## 6.　湯川はなぜ高知を訪ねたか：「ノーベル賞のもとは恩師三高校長森総之助」

　湯川が高知を訪れるにはそれなりの思いがあった。湯川は 1954 年 2 月 28 日京都で、高知訪問の仲介の労をとる高知三高同窓会「三高会」の二宮高知労働基準局長に会い次のように語っている。「恩師故森総之助元三高校長に物理の手ほどきを受けたので高知は特になつかしい、妻も高知行きを希望しているので出来れば同伴、龍河洞、桂浜などの景勝地を見物したい。」[52](湯川の高知訪問は同窓の大岡義明(高知赤十字病院長)が尽力し、1 月 13 日の時点で 3 月に来ることが決まっている。[53])

　湯川にとって三高校長であった森総之助とはどのような物理の先生であったのか。平沢興第 16 代京都大学総長(1900(明治 33)-1989 年(平成元年))は 1988 年出版の『現代の覚者たち』(p.190)で次のように述べている[54]。

　　「**平沢**　ノーベル賞は、よく京都から出るといわれますね。いままで日本で七人ノーベル賞もらっているんですが、そのうち五人、すなわち湯川秀樹、朝永振一郎、江崎玲於奈、福井謙一、利根川進氏は京都からです。そしてこのうち、初めの三氏はいずれも物理学の受賞者です。

　　これにはいろいろの原因があり、京都の自然とか、京大の個性を尊ぶ自由の空気なども関係しているでしょう。しかし、物理学賞の三氏などは、京都というよりも、京都の三高に物理学の先生で、森総之助という素晴らしい独創的な物理学の先生がおられたんです。この三氏は、森先生の教え子です。つまり、ノーベル賞のもとは森総之助なんです。

　　**Q**　その先生が偉かったんですか。

　　**平沢**　そう。この人は本当に学問好きな人。のちに、三高の校長になりましたが、こ

れは自分はなりたくないが周囲から推されて、やむなくなった。校長としても偉い人だったが、まぁ、物理学の先生としては本当に学問を愛した人なんですね。

　で、湯川君なんかは頭の回転の早い、いわゆる頭のいい人じゃなかったようです。大体、頭がいいとか悪いということが、実際どういうことなのか私にもよくわかりませんが、湯川君なんかは、すぐわかったような気持ちになる粗末な頭ではなく、わかるまで徹底的に考えぬく、限りない深さをもった頭ですね。

　平沢　だから、三日も四日も考えて、「どうも先生のいわれていることがわかっているようで、わからん」と質問に行くわけですね。そうすると、怠け者の先生は、「湯川、そんなことまで知らんでもいい。わしが教えたところまでわかれば、それでいい」とこういわれる。森先生はそうじゃない。先生は「そうか、湯川、お前はそこまで考えたか。たいしたもんだぞ」と。「それはいま世界の物理学界で問題になっているところだ。そこまで考えるとは偉いなぁ。お前はわしよりも偉いぞ」と、お世辞ではなく、心の底から感心をし、感心をしながらさらに話を進められるのです。そういう態度です。

　それは朝永君に対してもそうです。江崎さんは"また弟子"だそうです。そういうことで、学問に対する本当の教育というか、情熱を吹き込んだわけですね。

　Q　つまり、火をつけたわけですね。

　平沢　そう、火をつけた。教育とは火をつけることだ。教育とは、火をつけて燃やすことだ。教えを受けるとは、燃やされることであり、火をつけられることです。」

（下線は筆者）

　森総之助(1876(明治9)年5月11日-1953(昭和28)年4月23日)(図17)は高知県野市町出身の物理学者・理学博士で第6代第三高等学校の校長を務めた名教授であった。出身地の高知県野市町が編纂した『野市町史』[55][56]および『高知県人名事典』[57]、『土佐の科学技術者群像』[58]などによると、1876(明治9)年5月11日、高知県香美郡野市村東野19番屋敷で森為義の次男に生まれる。明治25(1892)年高知県立尋常中学校(現在の高知追手前高校)卒業後、京都の第三高等学校に入ったが、学制改変に伴い廃校になったので、級友の多くが東京の第一高等学校を選ぶのに対し、反骨の友人(高浜虚子、河東碧梧桐ら)と共に仙台第二高等学校に転配され入る。のち東京帝国大学理学部物理学科に入り、明治32(1899)年卒業。新潟県長岡中学校教諭を経て、明治34(1901)年母校第三高等学校教授となっている。

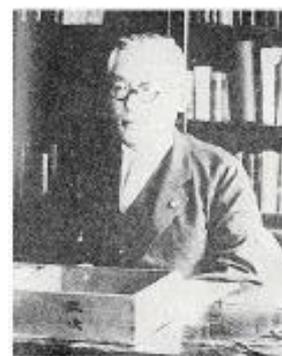

図17　湯川の恩師、第三高等学校校長、物理学者の森総之助(三高校長室)[56]。

　『野市町史』(下巻)[56]は「大正12年47歳にしてドイツに留学し3年間ミュンヘン大学ヴィーン博士(Wilhelm Wien、1864-1928年、明治44(1911)年ノーベル賞受賞)に師事して学識を高めたが、この一事をみても、彼の物理研究にかける情熱の一端を知ることができよう。」と記している[59]。

(森総之助の外遊中は 2 年生で「一戸・吉川両先生に、物理を習」い（[43]p.159）、3 年生で「学校でも力学を習うようになった」（[43]p.201）（力学担当は朝永振一郎の義兄堀健夫[60]）と湯川は記している。)『野市町史』（下巻）はさらに、「昭和 5 年から昭和 15 年の間、日本理化学協会副会長としてわが国物理教育界重鎮の地位にあったし、」溝淵進馬[61]第 5 代第三高等学校校長(高知県南国市大埆出身、1870-1935 年)の後任として「昭和 10 年第三高等学校校長となり、名校長の誉れが高かった」と記述している。

　『野市町史』（下巻）[56]は教育の具体的内容についても触れている。「彼は明治後期の新鋭物理学者・教育者として教育現場に臨んだが、その物理教育研究の特色とするところは、①実験を基本とした物理教育、②実践を土台とした物理教育研究、③<u>研究と教育を統一的にとらえる教育観</u>であった。当初は全般的に物理学に対する驚くべき基礎知識の貧困を痛感して、日本語版良教科書の作成と、不足する良教師の早期養成を第一義と考え、教師のための指導書編さんに意欲を燃やし、また多くの実験方法や実験器具の開発・製品化に努力して、実践を土台とした物理教育研究という理念を貫徹した。」さらに、「わが国における物理学教育の先覚者として活躍したが、明治・大正・昭和にかけて終始名利を求めず一学究をもって任じ、数多くの名著と実践を通じて、広くかつ深い感化を後進に与え、「森総」と通称されて、わが国物理学史上不朽の功績を残したのである。」（下線は筆者）

　平沢興『現代の覚者たち』[54]と同じく『野市町史』[56]も森総之助の第三高等学校での湯川秀樹、朝永振一郎(1906-1979 年)、江崎玲於奈(1925 年-)への物理教育に触れ「その<u>門下からは幾多の俊秀を輩出したが、ノーベル賞物理学者湯川秀樹・朝永振一郎また孫弟子として江崎玲於奈などが代表としてあげられよう。</u>」と記している（下線は筆者）。

　『野市町史』[56]はさらに次のように晩年までの大きい功績をたたえている。「多くの著書中主なるものをあげれば、『実践及ビ理論、物理学』『中学物理学教科書』『力学』『最新物理学　精義』などがある。また彼は昭和 15 年暮れ、にわかに」脳溢血のため「半身の自由を失い、よく 16 年三高校長を退職したが、頭脳および右手は健全であったので、病床にありながら日進月歩の物理学と取り組み、『新制高校物理解説』（昭和 24 年）を著している」。著書はほかにも、『最新物理学講義』（図18）、『実験及ビ理論物理学、音響学』、『実験及ビ理論物理学、光学』、『実験及ビ理論物理学、重学及物性論』、『実験及ビ理論物理学、電気及磁気学』（積善館、1917）、『実験及ビ理論物理学、熱学』、『実験及ビ理論、電気学』、『女子物理新教科書』（1924）、『物理解説　新制・高校　力学編』、『物理

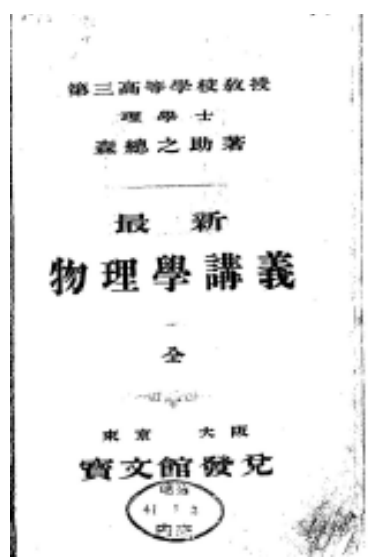

図 18　森総之助の執筆の本「最新物理学講義」（宝文館　1908 年発行）
（国立国会図書館蔵）

学講義実験法』、『新制中等物理学教科書』(1921)、『修正実業物理教科書』(1924)、『工業理科新物理』(1940)など多数ある。

　昭和28年4月23日午後5時京都市左京区北白川上池田町の自宅で死去。享年78歳(満76歳)。告別式は4月30日午後1時から京都黒谷三高会館で三高同窓会により行われ、「全国から参列した教え子や市民であふれ盛況であったという。生前正四位(正三位とも)勲二等瑞宝章を受ける。」[56][62]なお、京都府乙訓郡向町在住の会社社長をしている子息(67歳)は『父の遺志』として野市町役場に寄付をした[63]。『野市町史』(下巻)[56]には、「昭和44年子息茂雄氏により東野(石家)にあった生家や土地を処分した際、父総之助郷土愛の遺志として、代価の一部100万円を野市町に寄付した。町ではその趣旨にそい図書約1500冊を購入、本人母校である野市小学校に「森文庫」を創設して、遺志を永久に伝えることにした。」とある。

　森総之助は世間的な名利をいっさいもとめず晩年まで学問・教育ひとすじであった。おどろくほどの広範囲にわたる数多くの理論および実験にわたる物理の本・教科書などの執筆からもその熱情の一端をうかがい知ることができる。「物理の手ほどきを受けた」若き湯川秀樹も「森総」で知られていた森総之助の物理の専門書や教科書で勉強したことであろう***。森総之助が希望したのでもないのに第6代校長になったのも、同郷の溝淵進馬が現職のまま急逝したので断り切れずなったのではないか。溝淵の生家のある南国市大埇と森総之助の生地である香南市野市は近い。ふたりとも同じ高知尋常中学校でまなび、第三高等学校に進んでいる。森総之助は溝淵より6歳下である。前任の第5代校長溝淵も学問・教育ひとすじであった。溝淵は昭和4年濱口雄幸内閣誕生のとき濱口雄幸首相(溝淵は中学からの同級生)から文部大臣に懇請されたが、「教育者として終わりたい」と断ったという[64]。名校長と慕われ、第四高等学校には溝淵進馬の胸像がある。森総之助の功績がこんにちあまり知られてないのは残念に思う。筆者は高知市の小学校(小高坂小学校)へ転校するまで、小学4年9月まで、わたしと父の生家のある南国市十市の学校(十市小学校)で学んだが、その地は溝淵進馬の生地に近く森総之助の生家も遠くない。香南市はわたしの母方の祖父母の生地でもある。本稿を執筆する動機のひとつはわたしも学んだ湯川秀樹先生がわたしの同郷の森総之助先生から「物理の手ほどきを受け」[65]、大きな影響を受けていたことを知り、(わたしも森総之助の孫弟子ということになる)その功績を思い起こしたいことでもある。湯川は「恩師故元三高校長に森総之助から物理の手ほどきを受けた」[52]と語っているが、「物理の手ほどきを受ける」という湯川の言い方は単に本を読んだり話を聞いた

---

　*** 湯川が第三高等学校に入るまでに森総之助は次の本を出版していて、湯川は自伝[43]で「森総之助という有名な物理の先生」と記し、これらの本で勉強した思われる。森総之助著『最新物理学講義』(宝文館,1908)、『実験及ビ理論物理学.熱学』(積善館,1916)、『実験及ビ理論物理学.音響学』(積善館,1917)、『実験及ビ理論物理学.光学』(積善館,1917)、『実験及ビ理論物理学.重学及物性論』(積善館,1917)、『実験及ビ理論物理学.電気及磁気学』(積善館,1917)、『最新物理学精義』(積善館,1920)；森総之助編『物理学:実験及ビ理論.熱学』(積善館,1909)、『物理学:実験及ビ理論.音響学』(積善館,1909)、『物理学:実験及ビ理論.光学』(積善館,1909)、『物理学:実験及ビ理論.電気学』(積善館,1909)、『物理学講義実験法』(丸善,1911)。

というのではなく、物理への興味・関心を掻き立て研究への強い志しを後押しするような強い影響の受け方を指すように思われる。これは平沢興の語っていること[54]と符合する。

　三高で学んだ作家の田宮虎彦(1911-1988年)は第5代溝淵進馬校長と第6代森総之助校長の思い出を「三高時代の恩師」[66]で次のように語っている。「溝淵先生が三高の校長になられた時(昭和5年9月)自宅に生徒を数人ずつ呼んで親しく話をされたのです。（森総之助という人もえらい方だったそうですね。）森総之助先生には保証教授になってもらったんです。保証教授というのは現在の補導教官に似た制度で、下宿を変わったり、図書館で本を借りたりするときは保証教授のハンコが必要でした。」

　湯川が米国での5年間の滞在から日本に帰国するのは森総之助の逝去から約3か月後の1954(昭和28)年7月16日である。森総之助の京都の自宅・京都市左京区北白川上池田町および葬儀の行われた京都市左京区黒谷は京都大学および湯川の住む左京区下鴨からは遠くない。湯川が高知を訪問するのは帰国後8か月後であり、「恩師故森総之助元三高校長に物理の手ほどきを受けたので高知は特になつかしい」[52]と語ったのは、葬儀にも参列できなかったことの思いがあったかもしれない。

　湯川は夜須小学校での胸像除幕式を3月22日午前11時半に終えると直ちに夜須町の西方にある赤岡町の県立城山高校での講演会に向かっている。城山高校講堂では12時半から足立校長のあいさつ、国吉PTA会長による紹介のあと、湯川博士が登壇し、聴衆約2000余を前に「私は幼い時から志を立ててこれを実行してきた。将来もこれで進む覚悟である、あながち偉人になるのがえらいのではない、皆様は志を立てて将来一つの研究に努力

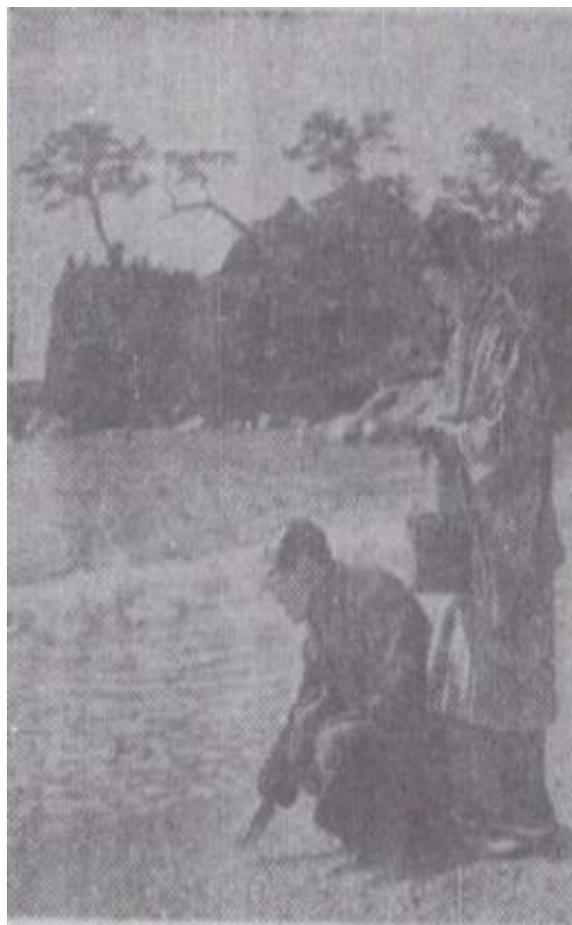

図19 景勝地桂浜を訪ねた湯川秀樹夫妻(1954年3月23日)[68]。

を傾け有能な人物になって世界人類のため尽くしてほしい」と約40分間にわたって講演している[67]。講堂に入りきれなかった聴衆のため高校玄関前で同じ話を行っている。

　講演終了後、澄子夫人とともに同高職員と記念撮影し、山口県の秋吉台とならぶ日本三大鍾乳洞の一つ、野市町に近い龍河洞の見物に向かっている。行く途中に通る野市町は赤

岡町に隣接し、野市東野には森総之助の生家があり野市西野には森総之助が学んだ野市小学校がある。湯川が森総之助の生家に立ち寄ったかは確認できてない。

　湯川が訪れた春分の日、3 月 22 日は気温 15 度のおだやかな天気で、野市には弘法大師・空海の四国八十八箇所霊場の第二十八番札所大日寺があり遍路の巡礼が始まる四国の春である。午後の龍河洞での息ぬきのあと夕方には高知市中央公民館での一般向けの講演会にのぞんでいる。湯川は高知訪問で胸像除幕式出席だけではなく、一般市民、学童生徒向けの一般講演、学術研究者にたいする専門的講演も行い、かなりきつい日程である。翌 3 月 23 日には午前に小・中学生のための講話、午後に学術講演を行い、3 時 58 分京都への列車で帰路についている。これらのことは別の稿で記すことにする。3 月 23 日は午前の講演が始まるまでの小時間を利用して、夫妻で希望していた桂浜を同窓生の案内で訪問している（図 19）[68]。雄大な太平洋をのぞむ坂本龍馬の銅像を見上げたことであろう。

## 7. 寺田寅彦銅像の設置運動－物理学者寺田寅彦と湯川秀樹

　私が恩師の湯川秀樹胸像に対面した同じ年、2018 年 7 月 24 日、高知市の旧追手前小学校跡地に新設開館の高知県市立図書館（オーテピア）に高知県出身の物理学者・寺田寅彦（1878（明治 11）－1935（昭和 10）年）の銅像（図 20）が設置され除幕式が行われた。銅像は彫刻家大野良一の制作で高知城をのぞむ追手筋に森総之助や寺田寅彦が卒業した高知追手前高校に面して設置されている。筆者も一員の「寺田寅彦記念館友の会」（山本健吉会長）を中心とした有志「寺田寅彦の銅像を建てる会」による 2014 年からの 1000 万円募金活動により建立された。銅像には寺田寅彦の有名な言葉「天災は忘れられたる頃来る」（台座左側面）と自筆短歌（台座右側面）とともに正面には「ねえ君　ふしぎだと思いませんか　寺田寅彦」と刻まれている。高知には偉大な物理学者の銅像がふたつあることになった。寺田寅彦像も湯川秀樹像とおなじく住民の自発的運動・募金で設置されたところに土佐の進取の民権自由思想が脈々と流れているように思われる。

　寺田寅彦は湯川の中間子論の論文[3]が出版された 1935 年に死去[†††]（12 月 31 日）しているが、理化学研究所で寺田寅彦の指導をうけた理論物理学者渡辺慧（1910-1993 年）によると湯川秀樹は理研時代の寺田寅彦と会っているようだ。湯川の記憶は鮮明でないようで[69]「私はどういうものか、寺田先生にはお目にかかった記憶はないね。一ぺんぐらいお会いしているんじゃないかと思うんだけども、記憶がありませんね。私がちょうど大阪大学におった時代で、そのころは東京の理研の仁科研究室にもよく出入りしておってね。」と渡辺との対談で語っている。東京帝国大学で物理の看板教授であった寺田寅彦は会議嫌い[‡‡‡]で、東京帝大に辞表を出し理化学研究所に移り（東京帝大は地震研究所所員で兼任）、1925 年から 1935 年晩年まで寺田研究室を主宰した。仁科芳雄（1890（明治 23）-1951（昭和 26）年）が理

---

[†††]寺田寅彦の墓は高知市東久万王子谷の寺田利正家墓地にある[70][71]。
[‡‡‡]　「『自分は何々の長と名のつくものには一生ならないつもりだ』[72]と宣言して、大学の管理運営や政治と一線を画した気骨のある人であった」[73]。

化学研究所に研究室を開くのは、ヨーロッパ留学（1921-1928年）から帰国してしばらくしてからの1931年で、寺田研究室よりだいぶ遅れる。仁科は1931年よく知られている京都大学での集中講義をおこない、湯川秀樹、朝永振一郎、坂田昌一らに最新のヨーロッパの原子核物理学を紹介している。湯川（寺田寅彦よりも28歳年下）は師の仁科芳雄や坂田昌一などのいる理化学研究所にはよく出かけており、寺田寅彦にあっている可能性は大きい。

　湯川秀樹が眺めたであろう桂浜の銅像の坂本龍馬は寺田寅彦と関係しているかもしれない。寺田寅彦の父・寺田利正は16歳のとき坂本龍馬[74]も関係するといわれる井口村刃傷事件§§§（文久元年（1861年）3月4日）で13歳の実弟・宇賀喜久馬の切腹を介錯したとされ[75]、寺田寅彦の遠戚の作家安岡章太郎(1920-2013年)が『流離譚』[75]で詳述している。父利正はトラウマになり、寺田寅彦も悲惨過ぎて直接的にかけず間接的に記述している[76]。

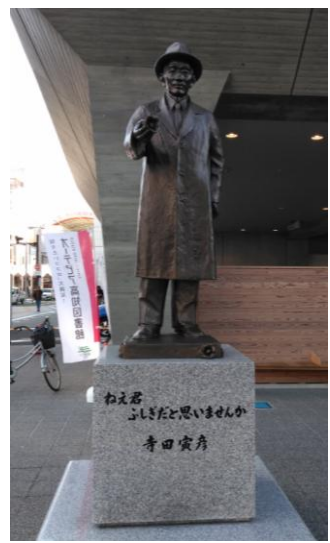

図20　2018年7月24日高知市城下に設立された寺田寅彦像。（2018年12月18日筆者撮影）

　寺田寅彦はドイツ・ベルリンに留学しプランク(M. Planck, 1858-1947年)の講義を聞き、英国ではラザフォードにマンチェスターで会っている[77]。寺田寅彦はラザフォードに会ってラジウムを見せてもらっている。ラザフォードが39歳で原子核を発見した論文を投稿した1911年である。ラザフォードの原子核発見の論文[17]の投稿は4月と論文中に記載されている（日は載っていない）ので、面会した1911年4月28日（金曜朝）は月末であり論文はすでに完成していたと考えられる。ラザフォードは論文の概要の内容を同年2月にManchester Literary and Philosophical Societyで発表しているので、寺田寅彦に会ったときには原子核の発見を確信している。（ラジウムの話だけでなく、原子の構造、原子核の発見について話が及んだと推察されるが、寺田の日記には記述がない。[77]）ラザフォードの原子核の発見が湯川秀樹の核力の研究、中間子論の発見[3]へと発展する。

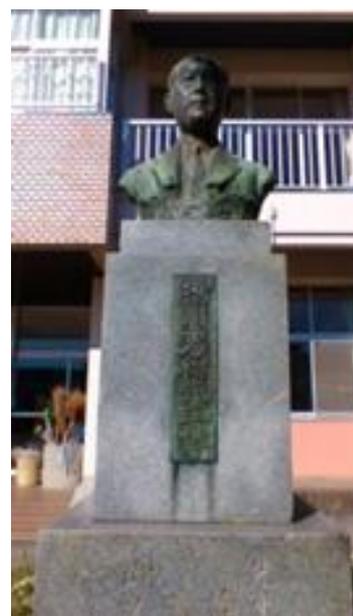

図21　高知県夜須小学校の「湯川秀樹先生像」の銘板と正面像。（2018年11月4日　筆者撮影）

---

§§§この事件を契機に5か月後8月に武市瑞山（武市半平太）（1829（文政12)-1865年（慶応元年））らによる土佐勤皇党結成で、討幕・明治維新へつながる。土佐勤皇党盟主の武市瑞山の銅像(原寛山作)は「武市半平太先生銅像建設期成会」により1979年高知県須崎市浦ノ内横浪半島に建てられている。

　寺田寅彦はヨーロッパに留学(1909(明治42)-1911(明治44)年)し、近代原子物理の最先端に触れ、1913(大正2)年「ラウエ斑点」でノーベル賞級といわれる研究を行いながら、(ラウエ(Max von Laue, 1879-1960年)は1914年ノーベル物理学賞、寺田は1917年帝国学士院恩賜賞受賞)ヨーロッパの要素還元主義的研究の後塵を追うことを嫌い独自の複雑系の物理を追究した。湯川(図21)と寺田(図20)、両者は一見きわめて対照的に見えるが、平沢興が「湯川君なんかは、すぐわかったような気持ちになる粗末な頭ではなく、わかるまで徹底的に考えぬく、限りない深さをもった頭ですね」という湯川と「ねえ君　ふしぎだと思いませんか」と複雑系の物理を徹底的に追究した(1908年の寺田の博士論文は「尺八の音響学的研究」)寺田とは根底において共通しているように思われる。寺田寅彦の教え子で物理学者・旧制高知高校教授の篠崎長之****は湯川が予言した新粒子にたいし、こんにち定着している「中間子」という名前を1939年に提唱していて[78]、不思議なつながり、因縁を感じさせる。寺田寅彦の家(現在の寺田寅彦記念館)は筆者が少年時代から過ごした小高坂で、坂本龍馬の家も近くであり、両人とも歴史上の人物であるとどうじに町内の先人という感じがする。大学勤務時代は高知城ちかくの寺田寅彦の家の前を歩いて通い「天災は忘れられたる頃来る」の碑を見ながらとくべつな親しみを感じたものである。湯川秀樹もその師森総之助も寺田寅彦も共通して晩年まで終生学問を愛し、多くの文筆・著書を残したが、歴史のなかでつながっているようで何か不思議な縁を感じさせられる。

## 8. おわりに

　高知県夜須町夜須小学校の湯川秀樹胸像は学校の正面玄関右側に設置され、建立当時は見晴らしの良いところで登下校の学童、教師、町民、誰でも目にはいった。いまは校舎は木造から鉄筋コンクリートに建て替えられ、胸像近くの松の木も大きく成長し、気をつけないと子どもたちには気づかれない。訪問時に校庭であそんでいる小学3年生の子どもに聞くと湯川像のことは知らなかった。学校・教育にかかわる大人にも尋ねてみたが、湯川像のみならず湯川秀樹の名前すら知らなかった。65年の時の変遷を考えると無理もないことかもしれない。教職員も子どもたちもみな過ぎ去った年月よりもはるかに若い。わたしは次の和歌を詠んで3か月後の短歌誌『塔』2月号に前田康子選歌で掲載された[30]。

　　その像はうまれる前から鎮座する教師子どもも知ることはなし

　夜須小学校の正面玄関の左側には湯川秀樹胸像と対をなして二宮尊徳の像がある(図2左)。背中に薪を背負い本を読んでいるよく知られた像である。夜須小学校に現在もこの像

---

　**** 1944年4月から清水高等商船学校(東京海洋大学の前身)へ転出。湯川が大阪大学で自由な気風の影響をうけた核物理学者菊池正士[79]は篠崎長之とともに寺田寅彦の実験指導をうけていて、寺田に提出した「学生実験レポート」の実物がみつかっている (高知県立歴史民俗資料館蔵)[71]。寺田寅彦が篠崎長之にあてた1935年9月25日付けの海鳴りに関する手紙が寺田の絶筆だと考えられている[71]。

があることに驚きを覚えつつも、楠公像にかわり、日本で最初に物理学者・湯川秀樹の胸像が私の生まれ故郷に作られ、永く保存されていることに感銘を受けた。湯川秀樹の恩師である森総之助の偉大さを知るにつれ、湯川秀樹が「物理の手ほどきを受けた」と語る恩師の生地・高知、香南市夜須小学校の校庭に「湯川秀樹胸像」が湯川秀樹夫妻の除幕で建立され、現在もあることは大変意義深いことだと思われた。

　湯川夫妻は1954年春の高知訪問の印象が強かったのか、1969年6月にふたたび高知を訪れている。62歳で翌年1970年3月31日の退官をひかえた年である。その1969年には、京都大学は大学紛争で荒れ狂い、それは湯川の基礎物理学研究所にも及んでいた。研究所の正面玄関には大きな字で中国の文化大革命でのスローガン「造反有理」という落書きがあった記憶が鮮明にある。この年は大学紛争で東大の入試が中止になり、京都大学の卒業式が中止になった記憶がある。私も大学の卒業式も大学院の入学式もないままいつとはなしに京都大学物理教室の大学院生になっていた。学内外が騒然としたなかで、湯川先生と基礎物理学研究所湯川記念館の正面玄関の前で卒業写真をとっていただいた（写真は[45]に掲載）。（このときは玄関前の湯川胸像はなかった。）湯川先生の直筆の時間・空間、物質・素粒子・素領域[80]にかかわる哲学的な深淵な意味を内包する[81]中国・盛唐の詩人・李白(701-762年)の詩[82]の色紙（図22）をいただいたのが昨日のように思い出される。

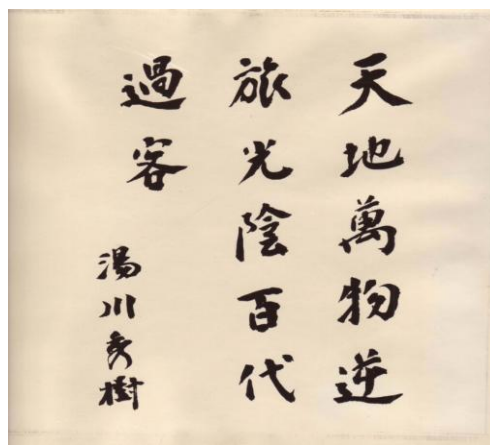

図22　湯川秀樹先生より京都大学卒業時いただいた色紙。(1969年3月)[45]

中国の古典、漢籍・漢詩などに造詣の深い[43][83]湯川先生は荘子の自然観を好まれ、荘子につながる思想をもった李白を好まれ[84]、李白のこの詩につながるよく知られた和歌がある[85]。

　　　　天地は逆旅なるかも鳥も人もいづこよりか来ていづこにか去る
　　　　（あめつち）（げきりょ）

　歌人紀貫之が土佐から京へ海路帰る途上、夜須町の手結（てい）海岸を太平洋から眺めその風光明媚を愛で和歌に残してからおよそ1000年、理論物理学者・歌人の湯川秀樹は京から鉄路高知へそして陸路夜須町を訪れた。紀貫之ははじめて仮名で日記をしるし文学にあたらしい世界を切りひらき、湯川秀樹は自然界に存在する4つの力(ニュートン(英)による重力、マクスウェル(英)による電磁気力、フェルミ(伊)[††††]による弱い力)、の4番目である「強い力」の核力を東洋人としてはじめて中間子理論で解明し、素粒子物理学を切り拓いた。

---

[††††] Enrico Fermi,1901-1954年

ニュートンの像は英ケンブリッジに、マクスウェルの像は英エディンバラにある。

　湯川秀樹先生の胸像建立と除幕式は町立夜須小学校 6 年生の卒業式にあわせ計画されたという。湯川秀樹はその日、第三高等学校時代に物理の手ほどき受けた恩師・森総之助の生地、高知のまちを訪ねた。1954 年 3 月 22 日、高知県夜須町の春は子どもたち・町民たち・除幕式出席の湯川秀樹夫妻にとって日ざしのやわらかい春分の日の、倖あふれしあたたかい「高知の春」、「四国の春」そして「日本の春」であった。湯川の随筆「四国の春」はのこされず、湯川胸像はその後語られることがなかった。

## 謝辞

## 参考文献

[1] 湯川秀樹、『しばしの幸』(読売新聞社、1954).

[2] 湯川秀樹、『湯川秀樹著作集　巻 7　回想・和歌』　p.222　(岩波書店、1989).

[3] H. Yukawa, 「On the Interaction of Elementary Particles」 Proc. Phys.-Math. Soc. Jpn. **17**, 48 (1935).

[4] 大久保茂男、「湯川秀樹先生の初めての一般講演」素粒子論研究電子版 **19**　(3) 2015 年 3 月
http://www2.yukawa.kyoto-u.ac.jp/~soken.editorial/sokendenshi/vol19/okubo.pdf、サイエンスポータル
https://scienceportal.jst.go.jp/news/newsflash_review/newsflash/2015/03/20150323_02.html

[5] 河辺六男、『湯川秀樹著作集別巻』(渡辺慧　編集) 年譜 (岩波書店、1990).

[6] S. Ohkubo and Y. Hirabayashi, 「Evidence for a secondary bow in Newton's zero-order nuclear rainbow」 Physical Review C (Rapid communications) **89** (5) 051601 (2014 年 5 月); 科学新聞「原子核に『2 番目の虹』存在」(2014 年 6 月 27 日); S. Ohkubo and Y. Hirabayashi,「Similarity between nuclear rainbow and meteorological rainbow: Evidence for nuclear ripples」 Physical Review C (Rapid communications) **89** (6) 061601 (2014 年 6 月); S. Ohkubo, Y. Hirabayashi, A. A. Ogloblin, Yu. A. Gloukhov, A. S. Dem'yanova, and W. H. Trzaska, 「Refractive effects and Airy structure in inelastic $^{16}$O+$^{12}$C rainbow scattering」 Physical Review C **90** (6) 064617

(2014年12月);大久保茂男 「原子核に副虹が存在」 パリティ 30 (2) 32-34 (2015年2月); R. S. Mackintosh, Y. Hirabayashi and S. Ohkubo,「Emergence of a secondary rainbow and the dynamical polarization potential for $^{16}$O on $^{12}$C at 330 MeV」Physical Review C 91 (2) 024616 (2015年2月); S. Ohkubo and Y. Hirabayashi,「Airy structure in $^{16}$O+$^{14}$C nuclear rainbow scattering」Physical Review C 92 (2) 024624 (2015年8月); S. Ohkubo, Y. Hirabayashi, and A. A. Ogloblin,「Further evidence for a dynamically generated secondary bow in $^{13}$C+$^{12}$C rainbow scattering」Physical Review C 92 (5) 051601(Rapid communications) (2015年11月); S. Ohkubo,「Luneburg-lens-like universal structural Pauli attraction in nucleus-nucleus interactions: Origin of emergence of cluster structures and nuclear rainbows」 Physical Review C 93 (4) 041303 (Rapid communications) (2016年4月); S. Ohkubo and Y. Hirabayashi,「Evidence for a dynamically refracted primary bow in weakly bound $^9$Be rainbow scattering from $^{16}$O」Physical Review C 94 (3) 034601 (2016年9月);S. Ohkubo, Y. Hirabayashi, and A. A. Ogloblin,「Existence of inelastic supernumerary nuclear rainbow in $^{16}$O + $^{12}$C scattering」 Physical Review C 96 (2) 024607 (2017年8月).

[7] 大久保茂男、「虹、核虹＝湯川虹からまなぶ理論物理学」素粒子論研究 117 (4) D119-121 (2009年10月).

[8] 湯川核力はふつう短距離(武谷三男の三段階理論の領域分けでの領域III)では強い斥力が働き「斥力芯」があると考えられているが成因は諸説あり現在でも完全にはわかっていない。パウリ原理から短距離で強い引力が働くと考えるのが「引力芯」説。両者は超対称性で理論的に等価と考えられている。S. Ohkubo、「Luneburg-lens-like structural Pauli attractive core of the nuclear force at short distances」Physical Review C 95 (4) 044002 (2017年4月).

[9] スピンが整数の湯川中間子や冷却原子ではボーズ・アインシュタイン凝縮が起こると考えられているが、スピン0のアルファ粒子でも同種の現象が原子核内で起こっているのではないかと考えられている。Y. Nakamura, J. Takahashi, Y. Yamanaka, and S. Ohkubo,「Effective field theory of Bose-Einstein condensation of α clusters and Nambu-Goldstone-Higgs states in $^{12}$C」Physical Review C 94 (1) 014314 (2016年7月); R. Katsuragi, Y. Kazama, J. Takahashi, Y. Nakamura, Y. Yamanaka, and S. Ohkubo,「Bose-Einstein condensation of α clusters and new soft mode in $^{12}$C-$^{52}$Fe 4N nuclei in a field-theoretical superfluid cluster model」Physical Review C 98 (4) 044303 (2018年10月).

[10]京都大学百年史編集委員会、『京都大学百年史：部局史編；3』p.180 (1997).

[11]京都大学百年史編集委員会、『京都大学百年史：部局史編；3』p.214 (1997).

[12]国際研究集会「QCDの新しい展開 2018」(5月28日－6月29)の期間中、国際会議「湯川国際セミナー2018b"クォーク・ハドロン科学の進展"」YKIS2018b Symposium 「Recent Developments in Quark-Hadron Sciences」(June 11-June 15, 2018)が開催され、筆者は次の核力の芯に関する論文を発表した。S. Ohkubo「Nuclear force with structural attractive core at short distances」http://www2.yukawa.kyoto-u.ac.jp/~nfqcd2018/YKIS/index.php

[13]「自然」増刊　追悼特集：「湯川秀樹博士『人と学問』」p.217（中央公論社、1981）.

[14]湯川秀樹、『天才の世界』（小学館、1973）.

[15]湯川秀樹、『続天才の世界』（小学館、1973）.

[16]湯川秀樹、『続々天才の世界』（小学館、1979）.

[17]E. Rutherford,「The scattering of α and β particles by matter and the structure of the atom」, Phil. Mag. **21**, 669-688 (1911).

[18]湯川秀樹、『湯川秀樹著作集　巻7　回想・和歌』（岩波書店、1989）.

[19]紀貫之、『土佐日記』（岩波書店、1979）；『続日本文学大系24』「土佐日記　蜻蛉日記　紫式部日記　更科日記」p.11　（岩波書店、1996）.

[20]湯川夫人の名前は1954年頃までの新聞等では湯川澄子という本名で記され、本稿では新聞記載事実に即し湯川澄子を使用する。現在知られている通称の湯川スミは1964年には使われており、朝日新聞1964年7月11日は「世界連邦建設同盟(会長、湯川スミ氏)」としている。以後も新聞等では通称が使われる。死去を伝える朝日新聞2006年5月15日は見出しで「湯川スミさん死去　９６歳　湯川博士の妻、核廃絶訴え」とし、本文記事で「日本人初のノーベル賞を受賞した物理学者の故湯川秀樹博士の妻で、世界連邦世界協会の名誉会長として核兵器の廃絶を目指す平和運動に尽力した湯川スミ（ゆかわ・すみ、本名・澄子＝すみこ）さんが、１４日午後２時４分、胃がんのため京都市左京区下鴨泉川町の自宅で死去した」と本名を記す。

[21]高知新聞社、『高知年鑑』（高知新聞社、1955）.

[22]吉川宏志は『塔事典』（永田和宏、花山多佳子、栗子京子監修、2014）などによると宮崎県東臼杵郡東郷町(現・日向市)出身(1969-)の歌人。歌集『青蝉』、『夜光』、『海雨』、『曳舟』、『西行の肺』、『燕麦』、『鳥の見しもの』、『石蓮花』がある。1987年京大短歌会。歌人前田康子は妻。第40回現代歌人協会賞受賞。

[23]小林幸子は『塔事典』（永田和宏、花山多佳子、栗子京子監修、2014）などによると奈良県宇陀郡室生村(現・宇陀市)出身(1945－)の歌人。前登志夫に師事。歌集に『夏の陽』、『枇杷のひかり』、『あまあづみ』、『千年紀』、『シラクーサ』、『六本辻』がある。

[24]短歌誌『塔』（京都市）2018年12月号　**65**巻12号　p.152-153.

[25]K. Ikeda, N. Takigawa, and H. Horiuchi, 「The Systematic Structure-Change into the Molecule-like Structures in the Self-Conjugate 4n Nuclei」Supplement of Progress of Theoretical Physics Extra Number p.464 (1968).この論文誌は小林稔京都大学教授の還暦記念の特別号として出版された。池田清美にとって小林稔は京都大学理学部小林研での師であり、池田はこの記念すべき論文を滞在中のソビエト連邦のDubna研究所で完成させたと記している。ノーベル物理学賞受賞者朝永振一郎の巻頭「Reminiscence」 https://doi.org/10.1143/PTPS.E68.3 や南部陽一郎の「Quantum Electrodynamics in Nonlinear Gauge」, https://doi.org/10.1143/PTPS.E68.190 　なども掲載されている。

[26]青木健一、坂東昌子、登谷美穂子編「素粒子論研究」**112**巻6号　(2006)『学問の系譜-アインシュタインから湯川・朝永へ-』池田清美、「原子核物理学の展開：はじめから終わりまでの繰り

返し」p.F14-F27；大久保茂男、「原子核物理学の展開：クラスター模型の展開」p.F28-F41；矢崎紘一、「原子核物理学の展開：クォーク模型と核力」p.F42-F46.

[27] 青木健一、坂東昌子、登谷美穂子編「素粒子論研究」112巻6号（2006）『学問の系譜-アインシュタインから湯川・朝永へ-』南部陽一郎、「基礎物理学の系譜：基礎物理学-過去と未来」p.F77-F91；九後太一、「素粒子論の未来へむけて：場の理論の発展と日本」p. F185-F198.

[28] 研究集会「Threshold rule 50」（10月3日―5日）で筆者はPart Ⅰで「池田図に魅せられて半世紀－池田図の実体化と普遍化をめざして」、Part Ⅱで最近の場の理論的クラスター模型研究[9]について包括的に紹介し、「超流動クラスター模型による軽い核・中重核におけるαクラスターのボーズアインシュタイン凝縮と南部ゴールドストーンソフトモード」について講演した。

[29] ルーマニアの研究者にドイツ・ケルン大学もふくむ共同研究に誘われ2018年12月に論文執筆開始、滞英中の3月に完成し米物理学会誌 Physical Review Cに投稿. P. Petkov, C. Müller-Gatermann, D. Werner, A. Dewald, A. Blazhev, D. Bucurescu, C. Fransen, J. Jolie, S.Ohkubo, and K.O. Zell, 「New lifetime measurements for the lowest quadrupole states in $^{20,22}$Ne and possible explanations of the high collectivity of the depopulating E2 transitions」.

[30] 短歌誌『塔』（京都市）2019年2月号66巻2号 p.113.

[31] 前田康子は『塔事典』（永田和宏、花山多佳子、栗子京子監修、2014）などによると兵庫県伊丹市出身(1966-)の歌人。歌集『ねむそうな木』、『キンノエノコロ』、『色水』、『黄あやめの頃』、『窓の匂い』がある。1987年京大短歌会。歌人吉川宏志は夫。

[32] 大久保茂男、徳島科学史研究会創立30周年記念講演「原子核発見100年をむかえて」徳島科学史雑誌（2012）No.30, p.3.

[33] 大久保茂男、「原子核発見100年をむかえて」素粒子論研究電子版 12（4）2012年8月
http://www2.yukawa.kyoto-u.ac.jp/~soken.editorial/sokendenshi/vol12/ohkubo.html.

[34] 浜口家は代々高知県夜須町手結(てい)浦に居住し土佐藩番役・老役を務めてきた名家で、浜口青果は『手結(てい)浦日抄』の著者浜口関衛門重孝より数えて6代目に当たり、明治18(1895)年2月1日浜口家の長男として父の勤務先の富山市で生まれる。3歳の時母を失い夜須町の手結(てい)の祖父母のもとで育つ[35]。

[35] 夜須町史編集委員会、『夜須町史』（上巻、下巻）（夜須町、1987）.

[36] 高知新聞、1954年2月6日（土曜日）朝刊.

[37] 高知新聞、1954年3月23日（水曜日）夕刊.

[38] 高知新聞、1953年12月6日（日曜日）朝刊.

[39] 西山庸平、『西山庸平著作集』（雄文閣、1931）.

[40] 野村邦近 監修、『普及版 吟詠教本 漢詩篇(三)』p.44（公益財団法人日本詩吟学院 2007）；漢詩原文 日柳燕石作『詠楠公』 「日東有聖人/ 其名曰楠公/誤生干戈世/提剣作英雄」;(通釈)日本に聖人がいて、その人の名を楠公と呼ぶ。彼は生まれたのが戦乱の世であったために、武器を持ち兵を率いて戦うことになり、その武名をあげて英雄となったのである。

[41] 野村邦近 監修、『普及版 吟詠教本 漢詩篇(二)』p.33（公益財団法人日本詩吟学院 2007）；漢詩原文 徳川斉昭作 『大楠公』「豹死留皮豈偶然/湊川遺跡水連天/人生有限名無尽/楠氏精忠万古伝」;(通釈)「豹は死して皮を留む、人は死して名を残す」というのは決して偶然のことではない。大楠公楠木正成が戦死をとげた湊川の遺跡は、その川の流れが天までつづいているように、いつまでも残るだろう。人生には限りがあるが、その人の残した功績は、尽きることはない。大楠公の、この上ない忠誠は永久に伝わって消えることがないだろう。

[42] 高知新聞、1954 年 3 月 22 日朝刊.

[43] 湯川秀樹、『旅人 ある物理学者の回想』（朝日新聞社、1958）.

[44] 高知新聞、1954 年 3 月 26 日(金曜日)夕刊.

[45] 大久保茂男、「湯川先生の思い出－生誕 100 年記念を迎えて」『ふまにすむす』18 巻（2007 年 3 月）p.29-32.

[46] 朝日新聞、1950 年 9 月 2 日朝刊.

[47] 川端康成、『舞姫』（角川文庫、1954）.

[48] 湯川秀樹、『湯川秀樹著作集 巻 7 回想・和歌』p.325（岩波書店、1989）.

[49] アブラハム・パイス(杉山滋郎・伊藤伸子訳)、『物理学者たちの 20 世紀 ボーア、アインシュタイン、オッペンハイマーの思い出』p.481（朝日新聞社、2004）.

[50] 朝日新聞、1953 年(昭和 28 年) 7 月 17 日(金曜日)朝刊.

[51] 湯川秀樹他監修、『理科 中学校 3 年 下』（学校図書株式会社、1954）.

[52] 高知新聞、1954 年 3 月 6 日(土曜日)朝刊.

[53] 高知新聞、1953 年 1 月 13 日 朝刊.

[54] 平沢興、『現代の覚者たち』p.185-211 （竹井出版、1988）.

[55] 野市町史編纂委員会、『野市町史』（上巻）(野市町、1992).

[56] 野市町史編纂委員会、『野市町史』（下巻）(野市町、1992).

[57] 高知県人名事典新版刊行委員会、『高知県人名事典 新版』（高知新聞社、1999）.

[58] 高知県技術者協会、『土佐の科学技術者群像』（高知県技術者協会、2003）.

[59] 湯川秀樹は『旅人 ある物理学者の回想』（朝日新聞社、1958）p.192[43]で森総之助の 47 歳での外遊について「三高には古くから、森総之助という有名な物理の先生がおられた。しかし、私が物理を習いはじめる時、この先生は外国へ行っていた。高校教授の外国出張は、珍しいことであった。」と記し、森総之助の研究への熱情を書き留めている。

[60] 日本物理学会、「物理学史資料委員会会報」No.4 p.9 （2011 年 3 月）.

[61] 高知市民図書館編、『高知市民図書館所蔵 特設文庫総合目録 下巻』「溝渕進馬資料」（高知市民図書館、1999 年)や[57]などによると溝淵進馬は明治 3(1870)年 12 月 25 日高知県長岡郡大埆村(南国市)の郷士・溝淵渉の 3 男に生まれ、明治 21(1888)年高知尋常中学校(現・高知追手前高校)を経て京都の第三高等学校へ進学。明治 28(1895)年帝国大学文科大学を卒業、母校高知尋常中学校の教諭となる。明治 30(1897)年仙台の第二高等学校教授、明治 31(1898)年千葉尋常中学校校長(1898-1899 年)。明治 32(1899)年から 3 年間ドイツ、フランスに留学しイエーナ大学、パ

リ大学で教育学をはじめ哲学、歴史、文学と幅広く学ぶ。帰国後、東京師範学校教授、東北帝国大学農科大学教授(北海道)、予科練教授(1908-1911年)を経て金沢の第四高等学校校長(1911-1921年)、熊本の第五高等学校教授校長(1921-1931年)を歴任。昭和6年第三高等学校校長に就任。昭和10(1935)年9月11日現職のまま京都で病没した。66歳。墓は南国市大埆にある。前身が第三高等学校の京都大学人間・環境学研究科総合人間学部図書館には溝淵文庫がある。

[62] 高知新聞、1953年4月25日．

[63] 高知新聞、1968年4月24日．

[64] 南国市史編纂委員会、『南国市史』（下）p.972（1982）．

[65] 森三高校長については第4代森外三郎校長と混同される向きが無いよう補足する。湯川が」京都一中に入学したときの校長は森外三郎であるが、後に三高に移るので、湯川が三高に入学したときの校長も森外三郎である[43]。[43]で述べられているように「温厚な紳士であると共に、古武士的な」森外三郎は数学者（著書に『代数学教科書』、『幾何学初歩』などがある）であり、湯川は物理の影響・手ほどきを受けてないことは言うまでもない。第6代校長の森総之助は平沢興が[54]湯川に影響を与えたと語り、また湯川が「恩師森総之助から物理の手ほどきを受けた」[52]と語っているが、具体的な記述は[43]には書かれてない。森総之助が3年間の外遊するのは大正12(1923)年からであり、湯川が高校に入学したのは同じ大正12(1923)年4月で3年間在学する。湯川は森総之助に「物理を習わなかった」とは[43]には書いていない、一方、3年で「学校でも力学を習うことになった」（[43]p.201）と記し、学校で習う前に自分で力学を勉強していたことをうかがわせる記述である。本を買うのが好きな湯川は三高の「有名な物理の先生」[43]である森総之助の本で勉強したのではないかと思われる。今後森総之助、湯川に関する新しい資料がでてくるか待ちたい。

[66] 田宮虎彦、高知新聞 1959年6月18日夕刊「南風対談第18回 田宮虎彦：三高時代の恩師」．

[67] 高知新聞、1954年3月23日（火曜日） 朝刊．

[68] 高知新聞、1954年3月24日 朝刊．

[69] 湯川秀樹、『湯川秀樹著作集別巻対談』p.301 「まとめるということ－寺田研とたどんの研究」（岩波書店、1990）．

[70] 高知県教育員会、『生誕百年記念増補改訂 寺田寅彦郷土随筆集』p.280．（1978年11月）．

[71] 上田壽、『寺田寅彦断章』「寺田家の墓地のことなど」p.201 （高知新聞社、1994）．

[72] 宇田道隆、「寺田寅彦先生の面影」寅彦研究 **15**、p.7－9（岩波書店、1937）．

[73] 酒井邦嘉、『科学者という仕事―独創性はどのように生まれるか』p.153 （中公新書、2006）．

[74] 坂崎紫瀾、『汗血千里の駒 坂本竜馬君之伝』（岩波書店、2010）．

[75] 安岡章太郎、『流離譚』（上） p.79 （新潮社、1981）．

[76] 安岡章太郎、『流離譚』（上） p.85-87 （新潮社、1981）．

[77] 寺田寅彦、『寺田寅彦全集』 第19巻 日記 2（岩波書店、1998）．

[78] 篠崎長之、「Heavy electronに對する譯語の一提案」科学 **9**（3）p.117（1939年3月号）（岩波書店、1939）．

[79] 湯川秀樹、『湯川秀樹著作集 巻7 回想・和歌』p.150 （岩波書店、1989）．
[80] 湯川秀樹、『湯川秀樹著作集 巻9 学術篇II』 p.261 （岩波書店、1989）の第9章で「素粒子と時空」で素領域の定式化があたえられており、§9.1は「ひろがりをもつ素粒子と素領域」である。湯川自身による素領域概念の説明は、湯川秀樹、『湯川秀樹著作集 巻3 物質と時空』「『素領域理論』とは何か」p.44 （岩波書店、1989）にある。
[81] 湯川秀樹、『湯川秀樹著作集 巻3 物質と時空』「天地万物逆旅」p.3 （岩波書店、1989）．
[82] 李白、「春夜宴桃李園一序」「**夫天地者萬物之逆旅** 光陰者百代之過客/ 而浮生若夢 爲歡幾何/古人秉燭夜遊 良有以也/ 況陽春召我以煙景 大塊假我以文章/ 會桃李之芳園 序天倫之樂事/群季俊秀 皆爲恵連/ 吾人詠歌 独慚康楽/ 幽賞未已 高談転清/ 開瓊筵以坐花 飛羽觴而酔月/ 不有佳作 何伸雅懐/ 如詩不成 罰依金谷酒数」 （星川清孝 『古文真宝（後集）』（新釈漢文大系 16）（明治書院 、1963）．
[83] 湯川秀樹、『湯川秀樹著作集 巻6 読書と思索』I.「東洋の思想」（岩波書店、1989）．
[84] 湯川秀樹、『続々天才の世界』p.8.「荘子」（小学館、1979）．
[85] 湯川秀樹、『湯川秀樹著作集 巻7 回想・和歌』p.359（岩波書店、1989）．

（2019年4月20日）

i ohkubo@rcnp.osaka-u.ac.jp